\documentclass[onecolumn,tighten,times]{aastex63}

\usepackage{amsfonts}
\usepackage{mathrsfs}
\usepackage{bm}
\usepackage{lipsum}
\usepackage{physics}
\usepackage{multirow}
\usepackage{rotating}
\usepackage{graphicx} 
\usepackage{soul}
\usepackage{subfigure}
\usepackage{color}

\def\bhm{M_{\bullet}}

\def\cblue{\color{black}}

\def\fBLR{f_{\rm BLR}}

\def\kms{\rm km\,s^{-1}}
\def\mathdotM{\dot{\mathscr{M}}}

\def\OIII{[O\,{\sc iii}]}

\def\RBLR{R_{\rm H\beta}}
\def\RL{R_{\rm H\beta}-L_{5100}}

\def\sigmad{\sigma_{\rm d}}
\def\sunm{M_{\odot}}
\def\taud{\tau_{\rm d}}
\def\pp{\prime\prime}

\newcommand{\feii}{\ifmmode {\rm Fe\ II} \else Fe~{\sc ii}\fi}
\newcommand{\heii}{\ifmmode {\rm He\ II} \else He~{\sc ii}\fi}
\newcommand{\oiii}{\ifmmode {\rm~[O\ III]} \else [O {\sc iii}]\fi}
\newcommand{\Rfe}{\ifmmode {{\cal R}_{\rm Fe}} \else ${\cal R}_{\rm Fe}$\fi}

\def\ergs{\rm erg\,s^{-1}}
\def\ergscm{\rm erg\,s^{-1}\,cm^{-2}}

\shorttitle{Reverberation Mapping of Two Luminous Quasars}
\shortauthors{Li et al.}

\begin{document}

\title{\bf \large
Reverberation Mapping of Two Luminous Quasars: the Broad-line Region Structure\\ \vglue 0.2cm 
and Black Hole Mass}

\author{Sha-Sha Li}
\affiliation{Key Laboratory for Particle Astrophysics, Institute of High Energy Physics, Chinese Academy 
of Sciences, 19B Yuquan Road, Beijing 100049, People’s Republic of China}
\affiliation{University of Chinese Academy of Sciences, 19A Yuquan Road, Beijing 100049, People’s Republic 
of China}

\author{Sen Yang}
\affiliation{Key Laboratory for Particle Astrophysics, Institute of High Energy Physics, Chinese Academy 
of Sciences, 19B Yuquan Road, Beijing 100049, People’s Republic of China}
\affiliation{University of Chinese Academy of Sciences, 19A Yuquan Road, Beijing 100049, People’s Republic 
of China}

\author{Zi-Xu Yang}
\affiliation{Key Laboratory for Particle Astrophysics, Institute of High Energy Physics, Chinese Academy 
of Sciences, 19B Yuquan Road, Beijing 100049, People’s Republic of China}
\affiliation{University of Chinese Academy of Sciences, 19A Yuquan Road, Beijing 100049, People’s Republic 
of China}

\author{Yong-Jie Chen}
\affiliation{Key Laboratory for Particle Astrophysics, Institute of High Energy Physics, Chinese Academy 
of Sciences, 19B Yuquan Road, Beijing 100049, People’s Republic of China}
\affiliation{University of Chinese Academy of Sciences, 19A Yuquan Road, Beijing 100049, People’s Republic 
of China}

\author[0000-0003-4042-7191]{Yu-Yang Songsheng}
\affiliation{Key Laboratory for Particle Astrophysics, Institute of High Energy Physics, Chinese Academy 
of Sciences, 19B Yuquan Road, Beijing 100049, People’s Republic of China}
\affiliation{University of Chinese Academy of Sciences, 19A Yuquan Road, Beijing 100049, People’s Republic 
of China}

\author{He-Zhen Liu}
\affiliation{School of Astronomy and Space Science, Nanjing University, Nanjing, Jiangsu 210093, People’s
Republic of China}
\affiliation{Key Laboratory of Modern Astronomy and Astrophysics(Nanjing University), Ministry of Education,
Nanjing, Jiangsu 210093, People’s Republic of China}

\author[0000-0002-5830-3544]{Pu Du}
\affiliation{Key Laboratory for Particle Astrophysics, Institute of High Energy Physics, Chinese Academy 
of Sciences, 19B Yuquan Road, Beijing 100049, People’s Republic of China}

\author[0000-0002-9036-0063]{Bin Luo}
\affiliation{School of Astronomy and Space Science, Nanjing University, Nanjing, Jiangsu 210093, People’s 
Republic of China}
\affiliation{Key Laboratory of Modern Astronomy and Astrophysics(Nanjing University), Ministry of Education,
Nanjing, Jiangsu 210093, People’s Republic of China}

\author{Zhe Yu}
\affiliation{Key Laboratory for Particle Astrophysics, Institute of High Energy Physics, Chinese Academy of
Sciences, 19B Yuquan Road, Beijing 100049, People’s Republic of China}
\affiliation{University of Chinese Academy of Sciences, 19A Yuquan Road, Beijing 100049, People’s Republic 
of China}

\author{Chen Hu}
\affiliation{Key Laboratory for Particle Astrophysics, Institute of High Energy Physics, Chinese Academy 
of Sciences, 19B Yuquan Road, Beijing 100049, People’s Republic of China}

\author[0000-0003-3825-0710]{Bo-Wei Jiang}
\affiliation{Key Laboratory for Particle Astrophysics, Institute of High Energy Physics, Chinese Academy 
of Sciences, 19B Yuquan Road, Beijing 100049, People’s Republic of China}
\affiliation{University of Chinese Academy of Sciences, 19A Yuquan Road, Beijing 100049, People’s Republic 
of China}

\author{Dong-Wei Bao}
\affiliation{Key Laboratory for Particle Astrophysics, Institute of High Energy Physics, Chinese Academy 
of Sciences, 19B Yuquan Road, Beijing 100049, People’s Republic of China}
\affiliation{University of Chinese Academy of Sciences, 19A Yuquan Road, Beijing 100049, People’s Republic 
of China}

\author{Wei-Jian Guo}
\affiliation{Key Laboratory for Particle Astrophysics, Institute of High Energy Physics, Chinese Academy 
of Sciences, 19B Yuquan Road, Beijing 100049, People’s Republic of China}
\affiliation{University of Chinese Academy of Sciences, 19A Yuquan Road, Beijing 100049, People’s Republic 
of China}

\author{Zhi-Xiang Zhang}
\affil{Department of Astronomy, Xiamen University, Xiamen, Fujian 361005, People’s Republic of China}

\author[0000-0001-5841-9179]{Yan-Rong Li}
\affiliation{Key Laboratory for Particle Astrophysics, Institute of High Energy Physics, Chinese Academy 
of Sciences, 19B Yuquan Road, Beijing 100049, People’s Republic of China}

\author{Ming Xiao}
\affiliation{Key Laboratory for Particle Astrophysics, Institute of High Energy Physics, Chinese Academy 
of Sciences, 19B Yuquan Road, Beijing 100049, People’s Republic of China}

\author[0000-0002-2310-0982]{Kai-Xing Lu}
\affil{Yunnan Observatories, Chinese Academy of Sciences, Kunming 650011, People’s Republic of China}

\author[0000-0001-6947-5846]{Luis C. Ho}
\affiliation{Kavli Institute for Astronomy and Astrophysics, Peking University, Beijing 100871, People’s 
Republic of China}
\affiliation{Department of Astronomy, School of Physics, Peking University, Beijing 100871, People’s 
Republic of China}

\author{Jing-Min Bai}
\affil{Yunnan Observatories, Chinese Academy of Sciences, Kunming 650011, People’s Republic of China}

\author[0000-0002-2121-8960]{Wei-Hao Bian}
\affiliation{Physics Department, Nanjing Normal University, Nanjing 210097, People’s Republic of China}

\author{Jes\'{u}s Aceituno}
\affiliation{Centro Astronomico Hispano Alem\'{a}n, Sierra de los filabres sn, E-04550 G\'{e}Tablesrgal, 
Almer\'{i}a, Spain}
\affiliation{Instituto de Astrof\'{i}sica de Andaluc\'{i}a (CSIC), Glorieta de la astronom\'{i}a sn, 
E-18008 Granada, Spain}

\author[0000-0002-2933-048X]{Takeo Minezaki}
\affiliation{Institute of Astronomy, School of Science, University of Tokyo, Mitaka, Tokyo 181-0015, Japan}

\author[0000-0003-1728-0304]{Keith Horne}
\affiliation{SUPA Physics and Astronomy, University of St. Andrews, North Haugh, KY16 9SS, UK}

\author[0000-0001-6402-1415]{Mitsuru Kokubo}
\affiliation{Department of Astrophysical Sciences, Princeton University, Princeton, New Jersey 08544,USA}

\author[0000-0001-9449-9268]{Jian-Min Wang}
\affiliation{Key Laboratory for Particle Astrophysics, Institute of High Energy Physics, Chinese Academy 
of Sciences, 19B Yuquan Road, Beijing 100049, People’s Republic of China}
\affil{University of Chinese Academy of Sciences, 19A Yuquan Road, Beijing 100049,  People’s Republic of China}
\affil{National Astronomical Observatories of China, Chinese Academy of Sciences, 20A Datun Road, Beijing 100020, People’s Republic of China}

\correspondingauthor{Jian-Min Wang, wangjm@ihep.ac.cn}

\begin{abstract}
We report the results of a multi-year spectroscopic and photometric monitoring campaign of two luminous 
quasars, PG~0923+201 and PG~1001+291, both located at the high-luminosity end of the broad-line region 
(BLR) size-luminosity relation with optical luminosities above $10^{45}~{\rm erg~s^{-1}}$.
PG~0923+201 is for the first time monitored, and PG~1001+291 was previously monitored but our campaign has 
a much longer temporal baseline. We detect time lags of variations of the broad H$\beta$, H$\gamma$, 
\feii\, lines with respect to those of the 5100~{\AA} continuum.
The velocity-resolved delay map of H$\beta$ in PG~0923+201 indicates a complicated structure with a mix
of Keplerian disk-like motion and outflow, and the map of H$\beta$ in PG~1001+291 shows a signature of 
Keplerian disk-like motion. Assuming a virial factor of $f_{\rm BLR}=1$ and FWHM line widths, we measure 
the black hole mass to be $118_{-16}^{+11}\times 10^7 M_{\odot}$ for PG~0923+201 and 
$3.33_{-0.54}^{+0.62}\times 10^7 M_{\odot}$ for PG~1001+291. Their respective accretion rates are estimated 
to be $0.21_{-0.07}^{+0.06} \times L_{\rm Edd}\,c^{-2}$ and $679_{-227}^{+259}\times L_{\rm Edd}\,c^{-2}$,
indicating that PG~0923+201 is a sub-Eddington accretor and PG~1001+291 is a super-Eddington accretor. 
While the H$\beta$ time lag of PG~0923+201 agrees with the size-luminosity relation, the time lag of 
PG~1001+291 shows a significant deviation, confirming that in high-luminosity AGN the BLR size depends 
on both luminosity and Eddington ratio. Black hole mass estimates from single AGN spectra will be 
over-estimated at high luminosities and redshifts if this effect is not taken into account.

\end{abstract}

\section{Introduction}
Over the past four decades, reverberation mapping (RM) technique 
\citep{Bahcall1972,Blandford1982,Peterson1993, Peterson2014} has been established as a standard tool for
measuring the mass of the central supermassive black hole (SMBH) and for studying geometry and dynamics of 
the broad-line region (BLR) in active galactic nuclei (AGNs). The characteristic time delay $\tau_{\rm BLR}$
between broad emission line and continuum variations corresponds to the light traveling time from the central
continuum source to the BLR, and therefore implies the BLR size by multiplying the time delay with the light
speed, namely, $R_{\rm BLR} = c\,\tau_{\rm BLR}$. Assuming the BLR is virialized, the black hole mass can be
determined through $R_{\rm BLR}$ and the broad emission-line width $\Delta V$,  
\begin{equation}\label{equ:bhm}
	\bhm = \fBLR\frac{R_{\rm BLR} \Delta V^2}{G},
\end{equation}
where $G$ is the gravitational constant and $\fBLR$ is the so-called virial factor depending on geometry 
and kinematics of the BLR (e.g., \citealt{Ho2014}). 

Up to now, there are about a hundred AGN RM observations and black hole mass measurements (e.g.,
\citealt{Peterson1998, Peterson2002, Peterson2004, Kaspi2000, Bentz2008, Bentz2009, Denney2009, Denney2010,
Barth2011, Barth2013, Barth2015,Grier2012, Grier2017, Du2014, Du2015, Du2016b, Du2018a, Du2018,Shen2016, 
Fausnaugh2017, De2018, Huang2019, Lu2019, Zhang2019, hu2020a, hu2020b,hu2021}). Those RM observations have
established the widely used $\RL$ relation between the size of H$\beta$ BLR ($R_{\rm H\beta}$) and the
monochromatic luminosity at 5100~\AA\, ($L_{5100}$) (e.g., \citealt{Kaspi2000,Peterson1999,Peterson2000,
Bentz2013}), which allows us to infer BLR sizes with single-epoch spectra and therefore to economically 
estimate black hole masses for large AGN samples. Needless to say, to reliably estimate black hole masses 
in single-epoch observations, the RM sample used for building up the $\RL$ relation needs to cover a variety 
of AGN populations. Recent RM campaigns indeed show that H$\beta$ time lags of super-Eddington accreting 
AGNs deviate from the $\RL$ relation \citep{Du2015,Du2016b,Du2018,Fonseca2020}.

On the other hand, due to the intensive time demands of RM observations, most of the previous RM campaigns 
were concentrated on low-luminosity and low-redshift sources. Time delays of those sources are relatively 
short (days or weeks) and the required monitoring periods are only several months. By contrast, RM of 
high-luminosity and high-redshift AGNs suffers the following restrictions. First, optical variability of 
high-luminosity AGNs is generally weak (e.g., \citealt{Hook1994, Giveon1999, Vanden2004, Kelly2009}), 
imposing difficulties on RM monitoring. Second, time delays for higher-redshift and higher-luminosity AGN 
are generally longer and monitoring periods are accordingly needed to be extended to years. Third, for 
high-redshift AGNs ($z\gtrsim0.7$), H$\beta$ emission lines are redshifted out of optical bands and 
infrared RM is required. Unfortunately, there has not been such infrared RM campaigns for H$\beta$ yet.
Hitherto, there are only a few RM sources at the high-luminosity ($L_{5100}>10^{45}~{\rm erg~s^{-1}}$) 
end of the $\RL$ relation. In this regard, it is quite necessary to expand the high-luminosity RM sample. 

To this end, we undertook a multi-year RM campaign on two luminous quasars, PG~0923+201 and PG~1001+291, 
both with an optical luminosity above $10^{45}~{\rm erg~s^{-1}}$. The obtained data allow us to measure 
the time delays of the broad H$\beta$, H$\gamma$, and \ion{Fe}{2} lines as well as to constrain the BLR
kinematics of the broad H$\beta$ line.

The paper is organized as follows. Observations and data reduction are described in 
Section~\ref{sec:observation}. In Section~\ref{sec:lc}, we explain in detail measurements of light 
curves of continuum and broad emission lines as well as intercalibrations of continuum light curves from 
our campaign and other time-domain survey archives. In Section~\ref{sec:result}, we present the time 
delay analysis and black hole mass measurements. In Section~\ref{sec:diss}, we discuss the implication 
of our results on the $\RL$ relation and the location of PG~0923+201 in the main sequence of RM samples. 
Here, the main sequence refers to the relation between full widths at half maximum (FWHMs) of broad 
H$\beta$ lines and strengths of \feii\, (\Rfe), where \Rfe\, is the flux ratio of \feii\, emission lines 
between 4434~\AA\, and 4684~\AA\, to broad H$\beta$ lines.
The conclusions are summarized in Section~\ref{sec:summary}. Throughout the paper, we use a $\Lambda$CDM
cosmology with $H_{0}=67\,\kms\,{\rm Mpc}^{-1}$, $\Omega_{\rm M}=0.32$, and 
$\Omega_{\Lambda}=0.68$ \citep{Planck2020}.

\begin{figure*}[th!]
\centering 
\includegraphics[width=0.45\textwidth]{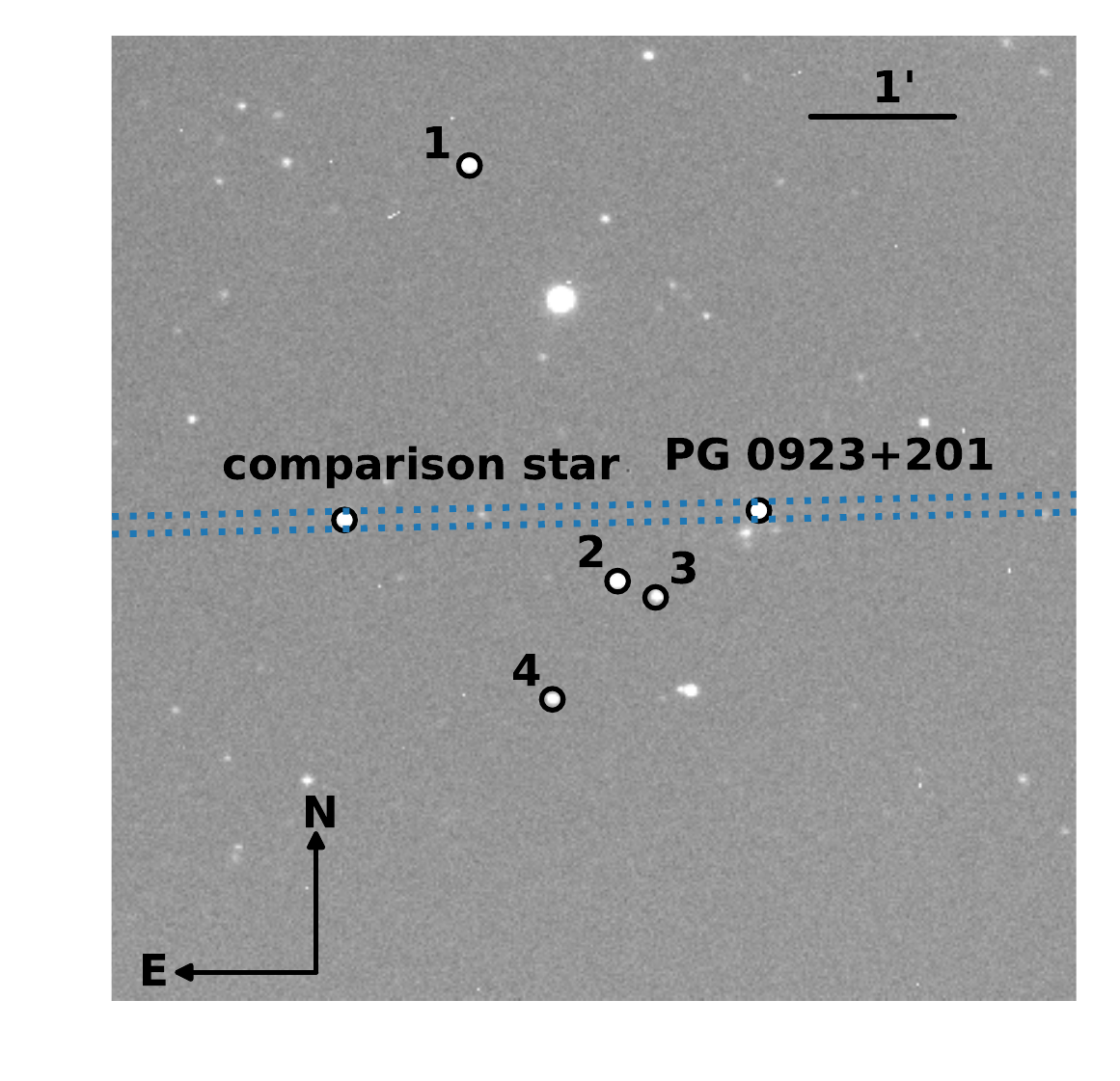}
\includegraphics[width=0.45\textwidth]{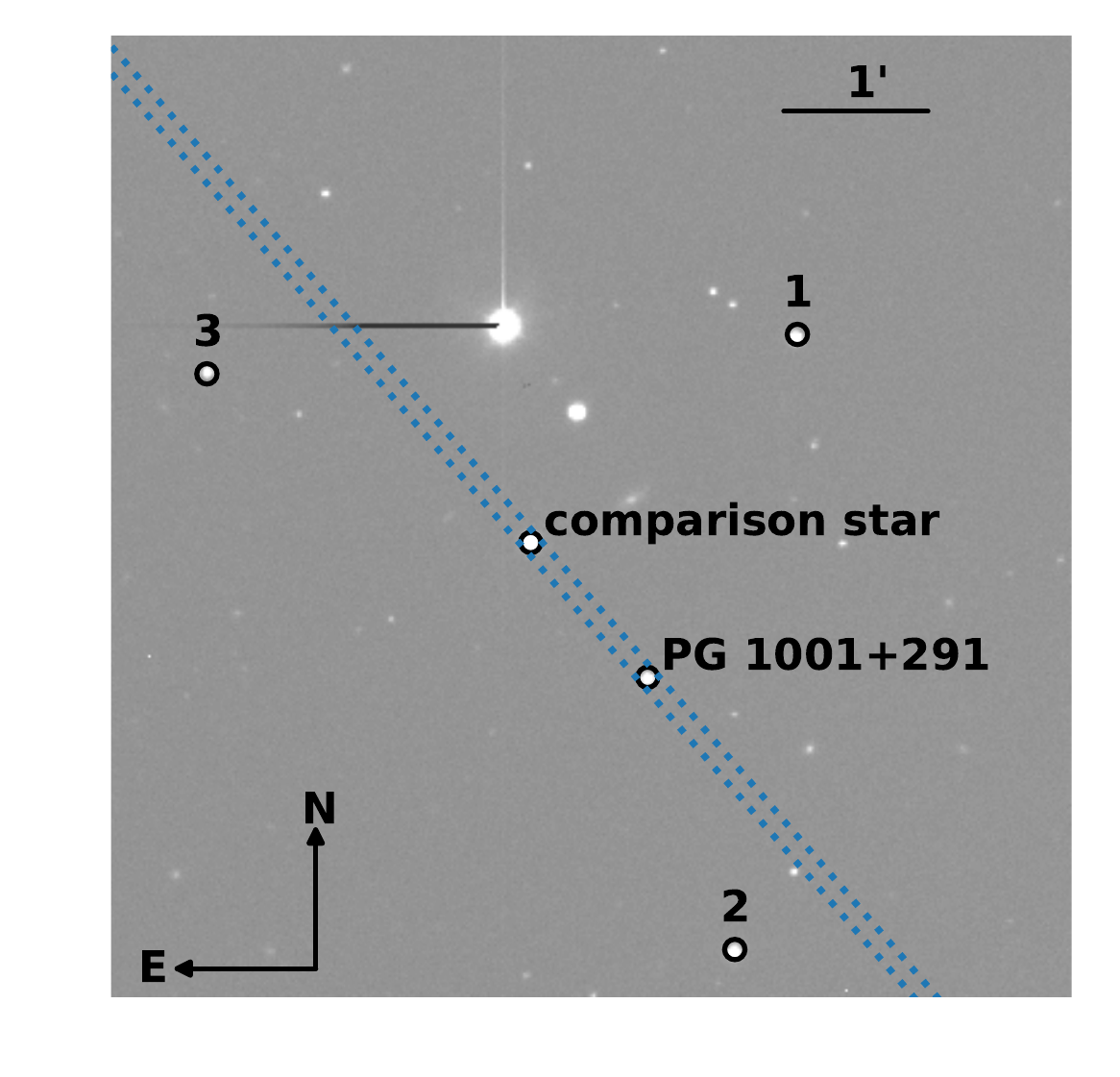}
\caption{Images of PG~0923+201 (left) and PG~1001+291 (right) and their comparison stars. Stars 1-4 in 
the image of PG~0923+201 and stars 1-3 in the image of PG~1001+291 are selected for differential photometry. 
The blue dotted lines illustrate the directions of slits.}
\label{fig:image}
\end{figure*}

\begin{figure}[th!]
\centerline{
\includegraphics[scale=0.60]{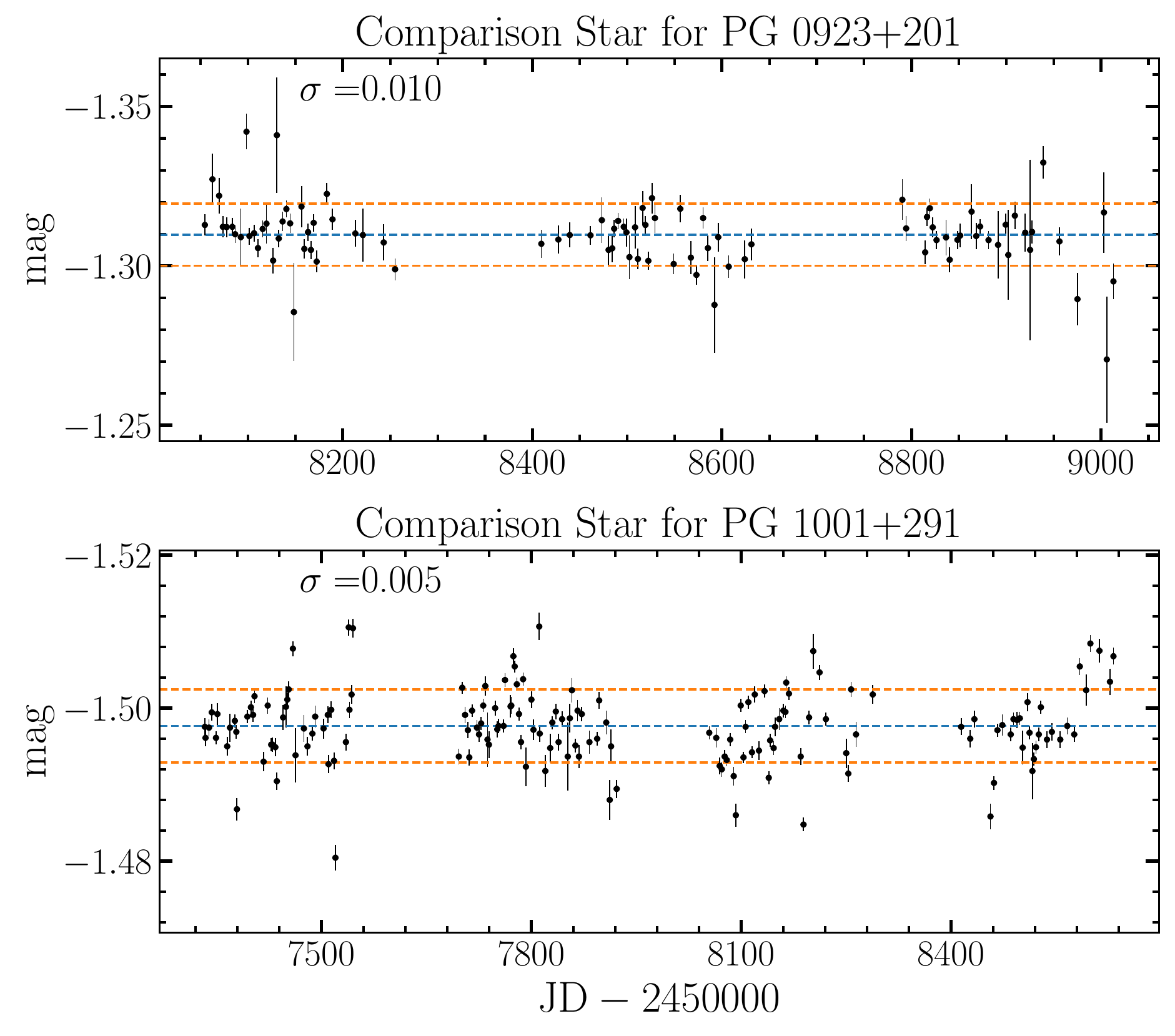}
}
\caption{Photometric light curves of the comparison stars for PG~0923+201 (top) and PG~1001+291 (bottom), 
in units of instrumental magnitudes. Blue dashed lines represent the mean magnitudes and orange dashed 
lines represent the standard deviations.}
\label{fig:comparestar}
\end{figure}

\begin{figure*}
\centering
\includegraphics[width=0.45\textwidth]{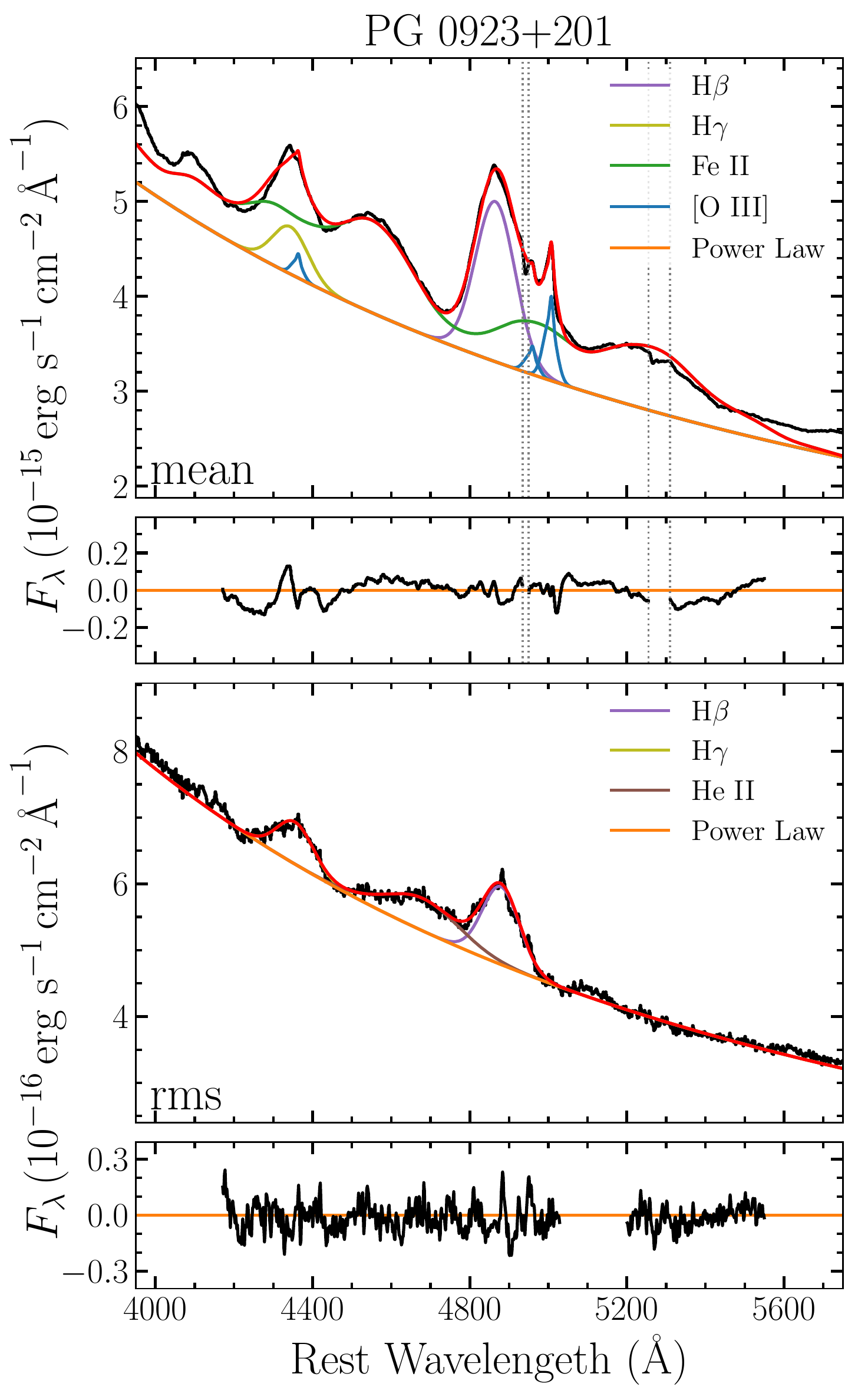}
\includegraphics[width=0.45\textwidth]{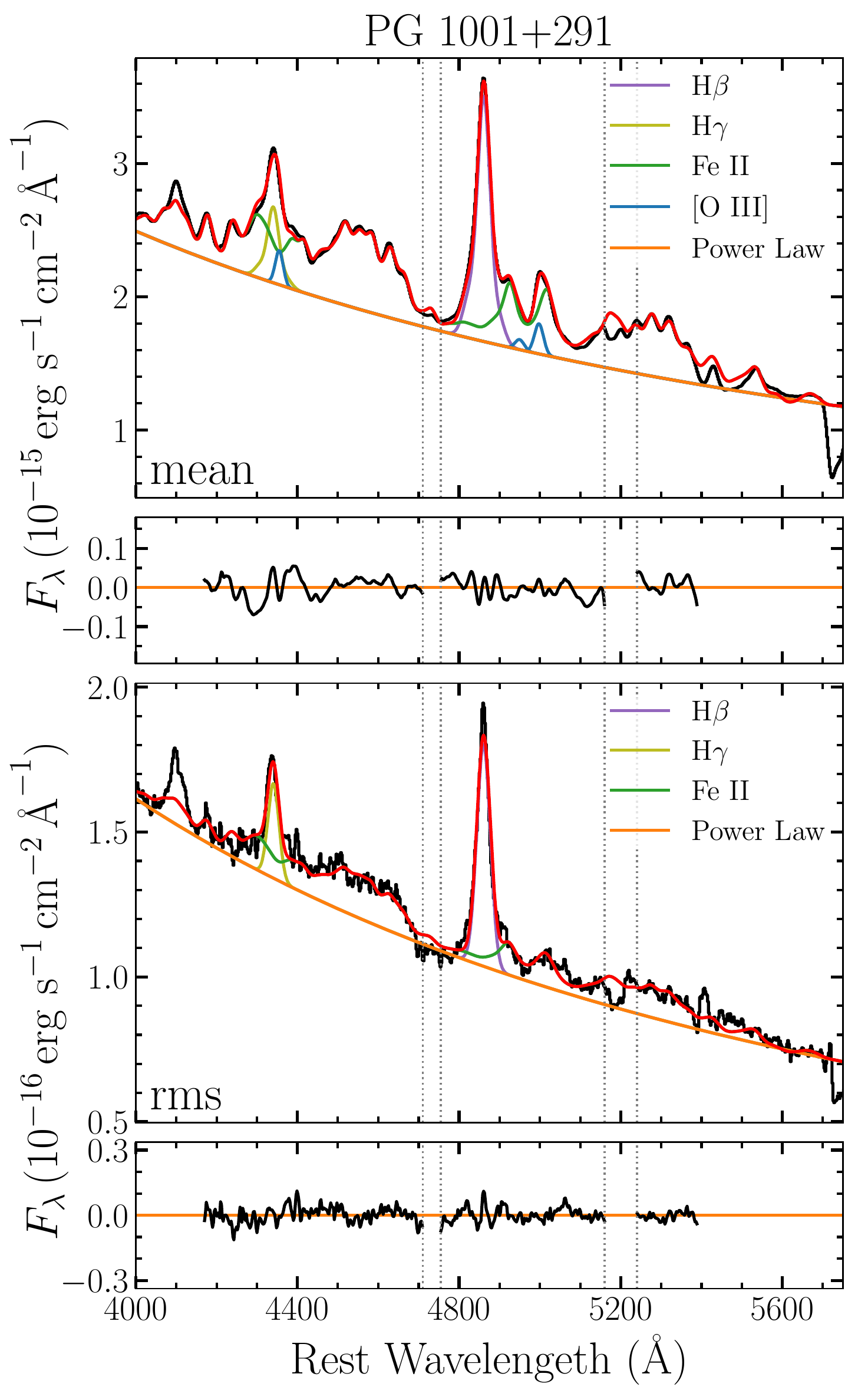}
\caption{Spectral decompositions of the mean spectrum (top) and the rms spectrum (bottom) for PG~0923+201 
in left panels and PG~1001+291 in right panels. In each case, two vertical adjacent panels show the details 
of spectral decompositions and the residuals in fitting windows. The vertical dotted lines indicate the 
regions omitted from the fits.}
\label{fig:fit}
\end{figure*}

\begin{deluxetable*}{lccccccc}
\renewcommand\arraystretch{1.1}
\tablecolumns{6}
\tablewidth{\textwidth}
\tablecaption{Properties of spectral and photometric data \label{table:obj}}
\tabletypesize{\footnotesize}
\tablehead{
\multirow{3}{*}{Object}                      &
\multirow{3}{*}{Source}           &
\multicolumn{2}{c}{Monitoring Period}           &
\colhead{}                         &
\multirow{3}{*}{Epochs}              &
\multirow{3}{*}{$\Delta T$ (day)}                     \\
\cline{3-4}
\colhead{}                         &
\colhead{}                         &
\colhead{JD - 2457000}                         &
\colhead{Date}                         &
\colhead{}                         &
\colhead{}                         &
}
\startdata
PG~0923+201              & LJS &1054 $-$ 1975      & 2017 Oct $-$ 2020 May && 89 & 5.9  \\
                         & LJP      &1054 $-$ 2013 & 2017 Oct $-$ 2020 Jun       && 85 & 6.0  \\
                         &ASAS-SN  &298 $-$ 2021  & 2015 Oct $-$ 2020 Jun  && 519 & 1.7  \\
                         &ZTF      &1202 $-$ 2000  & 2018 Mar $-$ 2020 May      && 374 & 0.1  \\
\hline 
PG~1001+291             & LJS &334 $-$ 1632      & 2015 Nov $-$ 2019 May && 164 & 4.9\\
                         & LJP      &333 $-$ 1632 & 2015 Nov $-$ 2019 Jun && 160 & 4.1\\
                        & ZTF      &1202 $-$ 1649  & 2018 Mar $-$ 2019 Jun && 143 & 1.0\\
                        & CAHAS    &894 $-$ 1492  & 2017 May $-$ 2019 Jan && 16 & 11.9\\
                        & CAHAP  &909 $-$ 1492    & 2017 Jun $-$ 2019 Jan && 14 & 11.1\\
\enddata
\tablecomments{LJS and CAHAS refer to spectroscopy from Lijiang and CAHA. LJP, CAHAP, ASAS-SN, and ZTF refer 
to photometry from Lijiang, CAHA, ASAS-SN archive, and ZTF archive, respectively. $\Delta T$ refers to the 
median sampling interval.
}
\end{deluxetable*}

\section{Observations and Data Reduction}\label{sec:observation}
\subsection{Targets}\label{sec:targets}
We spectroscopically monitored two quasars, PG~0923+201 and PG~1001+291, both with an optical luminosity 
larger than $10^{45}~{\rm erg~s^{-1}}$. Their basic properties are summarized below. 

PG~0923+201 has a redshift of $z=0.1921$, $m_{\rm g}=15.6~\rm mag$, and $V$-band extinction 
$A_{\rm V}=0.116\,\rm mag$ \citep{Schlafly2011}. 
Its spectrum shows prominent \feii\, emission lines, and weak \OIII\,\,$\lambda$5007 line \citep{Boroson1992}. 
The ratios of the equivalent width of \OIII\,\,$\lambda$5007 and \feii\, between 4434~\AA\, and 4684~\AA\, to 
that of H$\beta$ are about 0.04 and 0.72, respectively. Such strong \feii\, and weak \OIII\ emission lines are 
common features seen in narrow-line Seyfert~1 galaxies (NLS1s), which are generally believed to be accreting at 
a super-Eddington rate. However, the FWHM of the H$\beta$ line is as large as $\sim7600~\kms$, much broader than 
NLS1s (usually narrower than $2000~\kms$).

PG~1001+291 has a redshift of $z=0.3272$, $m_{\rm g}=15.9~\rm mag$, and $A_{\rm V}= 0.06~\rm mag$ 
\citep{Schlafly2011}. Similar to PG~0923+201, PG~1001+291 also shows prominent \feii\, emission lines  with 
\Rfe\,$\sim$\,1.17 and weak \OIII\ line \citep{Du2019}. However, its H$\beta$ line has a much narrower FWHM, only 
about $2100~\kms$. PG~1001+291 was monitored between 2015 and 2017 by the Super-Eddington Accreting Massive Black 
Holes (SEAMBH) campaign\footnote{The selection criteria for SEAMBHs were: strong optical \feii, relatively narrow 
H$\beta$, and weak \oiii\, lines (see \citealt{Du2014, Du2018}).} (\citealt{Du2018}, who used the alternative 
SDSS name for PG~1001+291, namely, SDSS~J100402.61+285535.3).
The H$\beta$ lag ($32.2_{-4.2}^{+43.5}$ days) was relatively uncertain due to the short monitoring period. 
Nevertheless, the H$\beta$ lag of PG~1001+291 was found to be 0.80~dex below the $\RL$ relation \citep{Du2018}.

\subsection{Spectroscopy} \label{sec:spec}
In our observations, spectroscopy was mainly undertaken with the Lijiang 2.4~m telescope at the Yunnan 
Observatories of Chinese Academy of Sciences. It is equipped with the Yunnan Faint Object Spectrograph and Camera 
(YFOSC) that can switch quickly between spectroscopy and photometry modes. The observations and data reduction 
were carried out following \cite{Du2014,Du2015}. We briefly summarize important points about observation setup 
below. (1) For PG~0923+201, the spectra were obtained by Grism~14 with a dispersion of 1.78~\AA~pixel$^{-1}$ and 
a wavelength coverage of $3800-7400$~\AA. We used a long slit with a projected width of $2.^{\pp}5$. The 
instrumental broadening was about 500~km~s$^{-1}$ (in terms of FWHM; \citealt{Du2014}). (2) For PG~1001+291, due 
to its larger redshift, Grism~14 would induce contamination of the second-order spectrum at the observed-frame 
wavelength longer than 6600~\AA\, (see \citealt{Lu2019,Feng2021}). We therefore used Grism~3 to avoid the 
contamination. Grim~3 has a spectral coverage of $3800-9000$~\AA\, and a dispersion of 2.9~\AA~pixel$^{-1}$. We 
adopted a $5.^{\pp}0$ wide slit and the instrumental broadening is about 1200~km~s$^{-1}$ \citep{Du2015}. 

For each target, we rotated the slit to cover a selected comparison star simultaneously for flux calibration 
(see Figure~\ref{fig:image}). {\cblue{We selected the comparison star WISEA~J092607.03+195357.5 for PG~0923+201 
and WISEA~J100406.45+285631.6 for PG~1001+291.}} The spectroscopic images were reduced using standard IRAF
procedures. A uniform aperture of $8.^{\pp}5$ and a background region of $7.^{\pp}4-14^{\pp}$ were used for 
spectrum extraction. Two consecutive 1200~s exposures were taken on each observing night. For PG~0923+201, we 
obtained a total of 89~epochs of spectroscopic observations between October 2017 and May 2020. For PG~1001+291, 
we obtained a total of 164~epochs between November 2015 and May 2019. {\cblue The median sampling intervals} of 
PG~0923+201 and PG~1001+291 are 5.9 and 4.9~days, respectively (see Table~\ref{table:obj}). For each night, the 
typical signal-to-noise ratio (S/N) per pixel at 5100~\AA\, (rest frame) is $\sim$ 57  for PG~0923+201 and $\sim$ 
48 for PG~1001+291.  

Since May 2017, PG~1001+291 was also observed with the Centro Astron\'omico Hispano-Alem\'an (CAHA) 2.2~m 
telescope at the Calar Alto Observatory in Spain (see \citealt{hu2020a, hu2020b}). The spectra were taken with 
the Calar Alto Faint Object Spectrograph (CAFOS) using the same observation strategy described above. The adopted 
Grism G-200 has a dispersion of 4.47~\AA~pixel$^{-1}$ and a spectral coverage of $4000-8500$~\AA. We used a long 
slit with a projected width of $3.^{\pp}0$ and the resulting instrumental broadening is about 1000~km~s$^{-1}$
\citep{hu2020b}. We selected the same comparison star as in our Lijiang observations. A uniform aperture of 
$10.^{\pp}6$ and a background region of $13.^{\pp}8-26.^{\pp}5$ were used for spectrum extraction (see 
\citealt{hu2020b} for data reduction details). In total, we obtained 16 epochs between May 2017 and January 2019 
at the Calar Alto Observatory. The typical S/N per pixel at 5100~\AA\, (rest frame) is $\sim$ 68 of each night.

The \feii\ blends are relatively strong beneath the \OIII\ line and the \OIII\ $\lambda$5007 line is too weak 
to apply \oiii-based calibration \citep{van1992}. Therefore, we adopt the method based on comparison stars, 
which provides sufficiently accurate flux calibration \citep{Maoz1990, Du2018}. Because the target and its 
comparison star are observed simultaneously, their emitted lights travel along the same path and thereby
suffer the same seeing and atmospheric conditions.
This enables high-accuracy relative flux measurements even in relatively poor weather conditions.
The calibration steps are as follows. {\cblue{First, we use the spectrophotometric standard stars to calibrate 
the absolute spectra of the comparison star observed on several good-weather nights. We average these calibrated 
spectra of the comparison star to generate a fiducial spectrum.
Second, in each observation, we compare the comparison star's spectrum with the fiducial spectrum to get a 
sensitivity function, and use this sensitivity function to calibrate the target's spectrum. We note that the 
absolute spectrum of the comparison star is not important, because RM 
analysis only depend on relative flux variations. Below, we also illustrate that the selected comparison stars 
are stable in photometry and therefore suitable for flux calibration.}}
Figure~\ref{fig:image} plots images to illustrate sky locations of the two quasars and their comparison stars 
along with slits.

\subsection{Photometry}  \label{sec:phot}
Photometry was obtained directly through the image mode of YFOSC. We used the Johnson $V$ filter and SDSS 
$r^{\prime}$-band filter for PG~0923+201 and PG~1001+291, respectively. Typically we took three 50~s exposures 
for each object on each observing night. We used standard IRAF procedures to reduce the photometric data. Several 
stars were selected in the ${\rm 10'\times10'}$ field of view for differential photometry (see 
Figure~\ref{fig:image}). According to aperture tests, we found that a circular aperture with a radius of 
$4.^{\pp}5$ and $4.^{\pp}2$ provides the best photometry for the two objects, respectively. The inner and outer 
radius of the background was uniformly set to be $8.^{\pp}5-17^{\pp}$. 
Figure~\ref{fig:comparestar} shows the photometric light curves of the comparison stars over our monitoring 
periods, both of which have a standard deviation $\lesssim$ 0.01~mag, indicating that the comparison stars were 
stable enough for flux calibration. 
We finally obtained a total of 85 and 160~epochs of photometric observations for PG~0923+201 and PG~1001+291 with 
median sampling intervals of 6.0 and 4.1~days, respectively (see Table~\ref{table:obj}). 
Tables~\ref{table:lightcurve1} and \ref{table:lightcurve2} list photometric light curves of the two targets.

We also obtained photometry of PG~1001+291 using the CAFOS image mode with a Johnson $V$ filter at the Calar 
Alto Observatory. Typically, three 60~s exposures were taken on each observing night. The photometry was 
reduced using standard IRAF procedures with the same configurations as in our Lijiang observations. We 
secured 14 epochs between June 2017 and January 2019 at the Calar Alto Observatory.

Besides our observations, we compiled photometry from archived data of the All-Sky Automated Survey for 
Supernovae (ASAS-SN) and the Zwicky Transient Facility (ZTF).

The ASAS-SN\footnote{\url{http://www.astronomy.ohio-state.edu/asassn/index.shtml}} is a long-term project
designed to survey the whole visible sky every night down to about 17 magnitude to discover nearby supernovae 
and other transient sources \citep{Shappee2014,Kocha2017}. The ASAS-SN project is composed of multiple stations, 
each station containing a 14 cm Nikon telephoto lenses equipped with a thermo-electrically cooled CCD camera. The 
field of view of each camera is about 4.5~$\rm deg^{2}$, and the pixel scale is $8^{\pp}$. Observations generally 
used $V$ or $g$ band filters for three dithered 90s exposures. The photometric data were processed using IRAF 
apphot package with an aperture radius of $16^{\pp}$ and calibrated according to AAVSO Photometric All-Sky 
Survey \citep{Henden2012}. For PG~0923+201, there are 519~epochs with {\cblue a median sampling interval} of
1.7\,days from October 2015 to June 2020 (see Table~\ref{table:obj}). For PG~1001+291, the ASAS-SN data are 
not adopted because of the relatively poor data quality.

The ZTF\footnote{\url{https://www.ztf.caltech.edu}} is designed for transients and variables and uses the 
Palomar 48-inch Schmidt Telescope with a 47 square degree field of view and a 600 megapixel camera to scan 
the entire northern visible sky \citep{Bellm2019,Graham2019,Masci2019}. The ZTF provides two filters ZTF-$g$ 
and ZTF-$r$, with the median photometric depths down to 20.8 and 20.6~mag, respectively. We combine the light 
curves of ZTF-$g$ and ZTF-$r$ bands using the intercalibration method described in Section~\ref{sec:lightcurve}. 
There are 374~epochs with typically 3 observations per night between March 2018 and May 2020 for PG~0923+201 and 
143\,epochs with a {\cblue median sampling interval} of 1.0~days from March 2018 to June 2019 for PG~1001+291 
(see Table~\ref{table:obj}).

\begin{deluxetable*}{cccccccccc}
\renewcommand\arraystretch{1.1}
 \tablecolumns{9}
\tablewidth{\textwidth}
 \tablecaption{Light curves of PG~0923+201 \label{table:lightcurve1}}
  \tabletypesize{\footnotesize}
  \tablehead{
     \multicolumn{5}{c}{Spectra}        &
      \colhead{}                         &
      \multicolumn{3}{c}{Photometry}     \\ 
      \cline{1-5} \cline{7-9}  
      \colhead{JD - 2450000}                       &
      \colhead{$F_{\rm 5100}$}               &
      \colhead{$F_{\rm H\beta}$}         &
      \colhead{$F_{\rm H\gamma}$}               &
      \colhead{$\rm Obs.$}               &
      \colhead{}                         &
      \colhead{JD - 2450000}     &
      \colhead{mag}          &
      \colhead{$\rm Obs.$}             
      }
\startdata
8054.37 & $26.45 \pm 0.37$ & $21.46 \pm 0.33$ & $5.74 \pm 0.19$ &  L && 7298.12 & $15.457 \pm 0.086$ & A \\
8065.37 & $26.48 \pm 0.38$ & $21.84 \pm 0.36$ & $5.71 \pm 0.22$ &  L && 7299.12 & $15.397 \pm 0.078$ & A \\
8069.43 & $26.81 \pm 0.36$ & $20.86 \pm 0.31$ & $5.44 \pm 0.17$ &  L && 7308.11 & $15.416 \pm 0.054$ & A \\
8073.35 & $27.75 \pm 0.37$ & $21.12 \pm 0.32$ & $5.86 \pm 0.19$ &  L && 7310.11 & $15.464 \pm 0.062$ & A \\
8077.37 & $27.47 \pm 0.37$ & $20.92 \pm 0.32$ & $5.56 \pm 0.19$ &  L && 7311.11 & $15.482 \pm 0.058$ & A \\
\enddata
\tablecomments{The 5100~{\AA} continuum flux densities are in units of $10^{-16}~\ergscm$~\AA$^{-1}$, and 
the fluxes of emission lines are in units of $10^{-14}~\ergscm$. {\cblue{In the ``Obs.'' column, ``L'' refers 
to Lijiang, ``A'' refers to ASAS-SN, and ``Z'' refers to ZTF.}}
\\
(This table is available in its entirety in a machine-readable form in the online journal.)}
\end{deluxetable*}

\begin{deluxetable*}{ccccccccccc}
\renewcommand\arraystretch{1.1}
 \tablecolumns{10}
\tablewidth{\textwidth}
 \tablecaption{Light curves of PG~1001+291 \label{table:lightcurve2}}
  \tabletypesize{\footnotesize}
  \tablehead{
      \multicolumn{6}{c}{Spectra}        &
      \colhead{}                         &
      \multicolumn{3}{c}{Photometry}     \\ 
      \cline{1-6}\cline{8-10}   
      \colhead{JD - 2450000}                       &
      \colhead{$F_{\rm 5100}$}               &
      \colhead{$F_{\rm H\beta}$}         &
      \colhead{$F_{\rm H\gamma}$}               &
      \colhead{$F_{\rm Fe\ II}$}         &
      \colhead{$\rm Obs.$}      &  
      \colhead{}                         &
      \colhead{JD - 2450000}                       &
      \colhead{mag} &
     \colhead{$\rm Obs.$}
         }
\startdata
7334.41 & $15.32 \pm 0.11$ & $8.60 \pm 0.11$ & $2.36 \pm 0.10$ & $11.20 \pm 0.23$ & $\rm L$ & & 7333.42 & $15.944 \pm 0.003$ &$\rm L $ \\
7339.40 & $14.78 \pm 0.12$ & $8.65 \pm 0.13$ & $2.20 \pm 0.11$ & $11.38 \pm 0.27$ & $\rm L$ & & 7334.40 & $15.945 \pm 0.003$ &$\rm L $ \\
7343.34 & $14.70 \pm 0.13$ & $8.29 \pm 0.14$ & $2.34 \pm 0.12$ & $11.13 \pm 0.29$ & $\rm L$ & & 7339.38 & $15.956 \pm 0.003$ &$\rm L $ \\
7349.38 & $15.11 \pm 0.10$ & $8.49 \pm 0.07$ & $2.41 \pm 0.06$ & $10.71 \pm 0.15$ & $\rm L$ & & 7343.33 & $15.962 \pm 0.003$ &$\rm L $ \\
7351.42 & $14.96 \pm 0.11$ & $8.72 \pm 0.10$ & $2.31 \pm 0.08$ & $11.38 \pm 0.19$ & $\rm L$ & & 7349.37 & $15.970 \pm 0.002$ &$\rm L $ \\
\enddata
\tablecomments{The 5100~{\AA} continuum flux densities are in units of $10^{-16}~\ergscm$~\AA$^{-1}$, and the fluxes of emission lines are in units of $10^{-14}~\ergscm$.
{\cblue{In the ``Obs.'' column, ``L'' refers to Lijiang, ``C'' refers to CAHA, and ``Z'' refers to ZTF.}}\\(This table is available in its entirety in a machine-readable form in the online journal.)}
\end{deluxetable*}

\section{Measurements} \label{sec:lc}
\subsection{Light Curves} \label{sec:sfs}
We measure the 5100~\AA\, flux density and broad H$\beta$ line via a multi-component spectral fitting scheme
(SFS) following \cite{hu2015,hu2020a,hu2020b}. Before fitting, we correct the Galactic extinction using the
extinction law of \cite{Cardelli1989} with $R_{V}=3.1$. We employ the DASpec\footnote{DASpec is available at
{\url{ https://github.com/PuDu-Astro/DASpec}.}} software to perform multi-component spectral fitting. The 
software uses the Levenberg-Marquardt method to minimize the chi-square. The details of the fitting procedures 
for the two targets are slightly different, as described below.

\paragraph{PG~0923+201} There is no obvious stellar absorption feature in our spectra ({\cblue{see 
Figure~\ref{fig:fit}}}; near $\sim$ 4940 and 5280~\AA\, are telluric absorptions), so we do not add a host 
galaxy template in the fitting. The fitting components include: (1) a single power law, (2) an \feii\,template 
from \cite{Boroson1992}, (3) a single Gaussian for broad emission lines H$\beta$, H$\gamma$, (4) double 
Gaussians for \OIII~$\lambda\lambda$4959, 5007 and \OIII~$\lambda$4363 with the same tied velocity shifts 
and widths among the three line. The narrow H$\beta$ is not added because the flux of the narrow H$\beta$ is 
less than 2\% of the total H$\beta$ flux, and the narrow H$\beta$ flux is estimated by assuming that a typical
flux ratio of 0.1 between the narrow H$\beta$ and \oiii~$\lambda$5007 (e.g., \citealt{Veilleux1987, hu2015}). 

Because the H$\gamma$ is blended with \OIII~$ \lambda$4363, we fix the velocity width and shift of H$\gamma$ 
to those of H$\beta$. The flux ratio of \OIII~$\lambda$4363 relative to \OIII~$\lambda$5007 is fixed to 0.251
obtained by the best fit of the mean spectrum. The \OIII\ lines have asymmetric blue wings, which is seen more 
clearly in the high-resolution spectrum from the Sloan Digital Sky Survey (SDSS). We therefore apply two 
Gaussians to fit the \OIII\ lines, one of which is set to be narrower and the other is set to broader. The 
width and shift of the broader Gaussian component of \OIII\ lines are fixed to the value obtained via the best 
fit to the SDSS spectrum after considering the different instrumental broadening of SDSS and Lijiang telescope. 
The fitting is performed at 4170-5550~\AA\, excluding the telluric absorption windows around 4940~\AA\, and 
5280~\AA. 

We note that although the \heii\,$\lambda4686$ line is relatively prominent in the root mean square (rms)
spectrum, it is weak and highly blended with the \feii\, and the blue wing of the H$\beta$ in individual 
spectra so that it cannot be well constrained in the SFS. We therefore do not include a \heii\, component. 
To evaluate the influences of the \heii\, line on the time delay measurements, we present the results by 
adding a Gaussian to account for the \heii\, component in Appendix~\ref{sec:add-heii-lc}. 
The obtained time delays from the two schemes are consistent within uncertainties, indicating that the 
influences of the \heii\, line are minor.

\paragraph{PG~1001+291} Again, the host galaxy template is not included. The fitting components include: 
(1) a single power law, (2) a \feii\ template from \cite{Boroson1992}, (3) a fourth-order Gauss-Hermite 
function for broad H$\beta$ and broad H$\gamma$, (4) a single Gaussian with the same velocity width and 
shift for narrow lines \OIII\,\,$\lambda\lambda$4959, 5007 and \OIII\,\,$\lambda$4363. Similar to PG~0923+201,
the width and shift of H$\gamma$ are fixed to those of H$\beta$, and the flux ratio of \OIII\,\,$\lambda$4363 
to \OIII\,\,$\lambda$5007 is fixed to the fitted value of the mean spectrum. The fitting is performed at
4170-5390~\AA, excluding the telluric absorption windows around 4730~\AA\, and 5200~{\AA} (see
 Figure~\ref{fig:fit}). 

The light curves of the continuum flux density at 5100~\AA\,($F_{\rm 5100}$) and the broad emission lines 
(H$\beta$ and H$\gamma$ of PG~0923+201; H$\beta$, H$\gamma$, and \feii\ of PG~1001+291) are directly obtained 
from the above fitting and summarized in Tables~\ref{table:lightcurve1} and \ref{table:lightcurve2}. The 
reported errors of the light curves include the Poisson errors and additional systematic errors (added in
quadrature). The additional systematic errors are determined following the procedure described in \cite{Du2014}.
We note that the current \feii\, data of PG~0923+201 is not enough to detect a reliable time delay so that 
we do not present \feii\ analysis for PG~0923+201 in this work.

\begin{figure}[t!]
\centerline{
\includegraphics[scale=0.6]{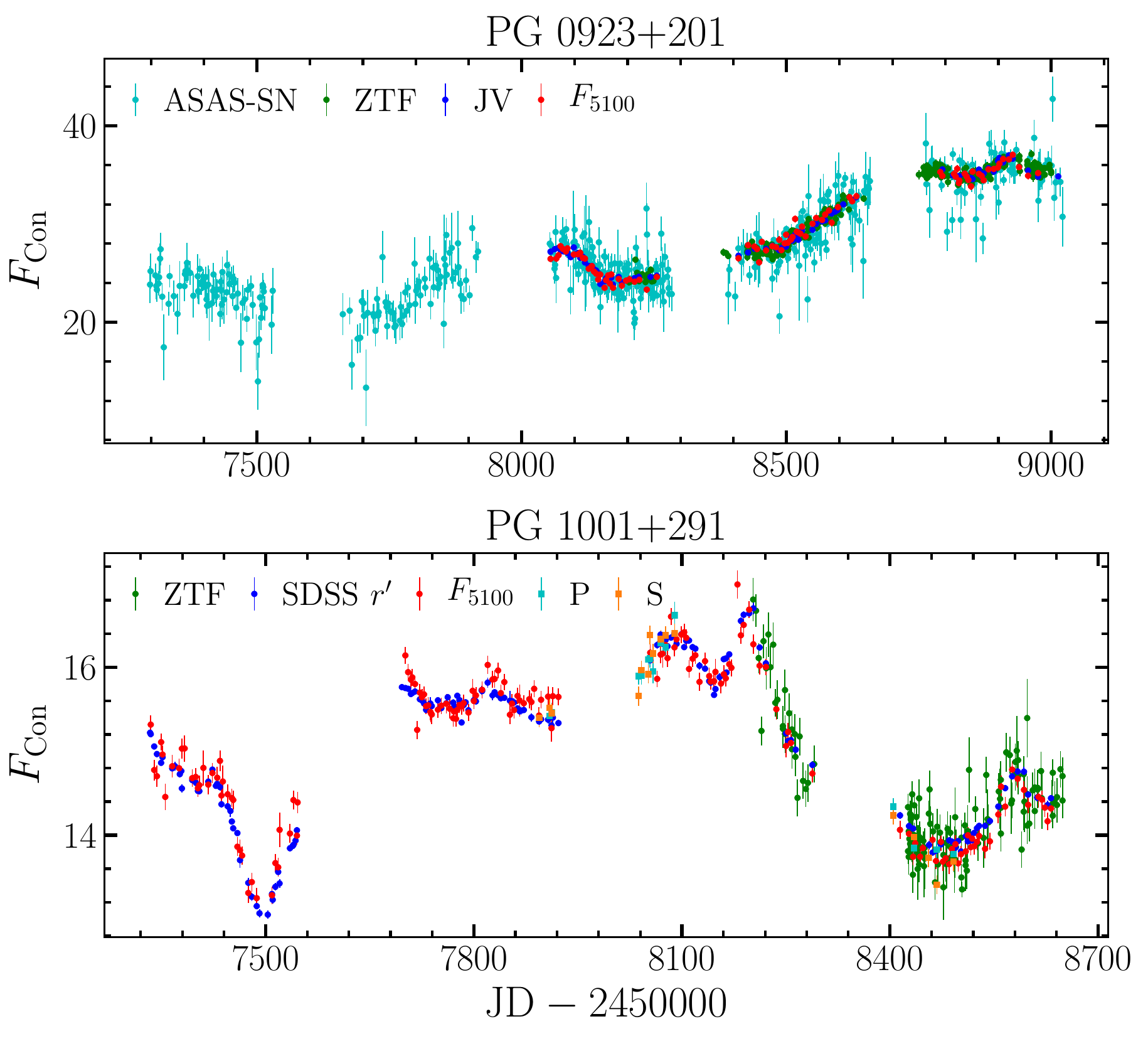}
}
\caption{The intercalibrated continuum light curves that combine the 5100~\AA\, continuum ($F_{\rm 5100}$), 
and photometry from our Lijiang observations (JV/SDSS $r^{\prime}$) and ASAS-SN and ZTF archives. ``P'' and 
``S'' refer to photometric data and 5100~\AA\, continuum from CAHA observations, respectively.}
\label{fig:intercali}
\end{figure}

\subsection{Intercalibration}\label{sec:lightcurve}
\begin{figure*}
\centering
\includegraphics[width=0.8\textwidth]{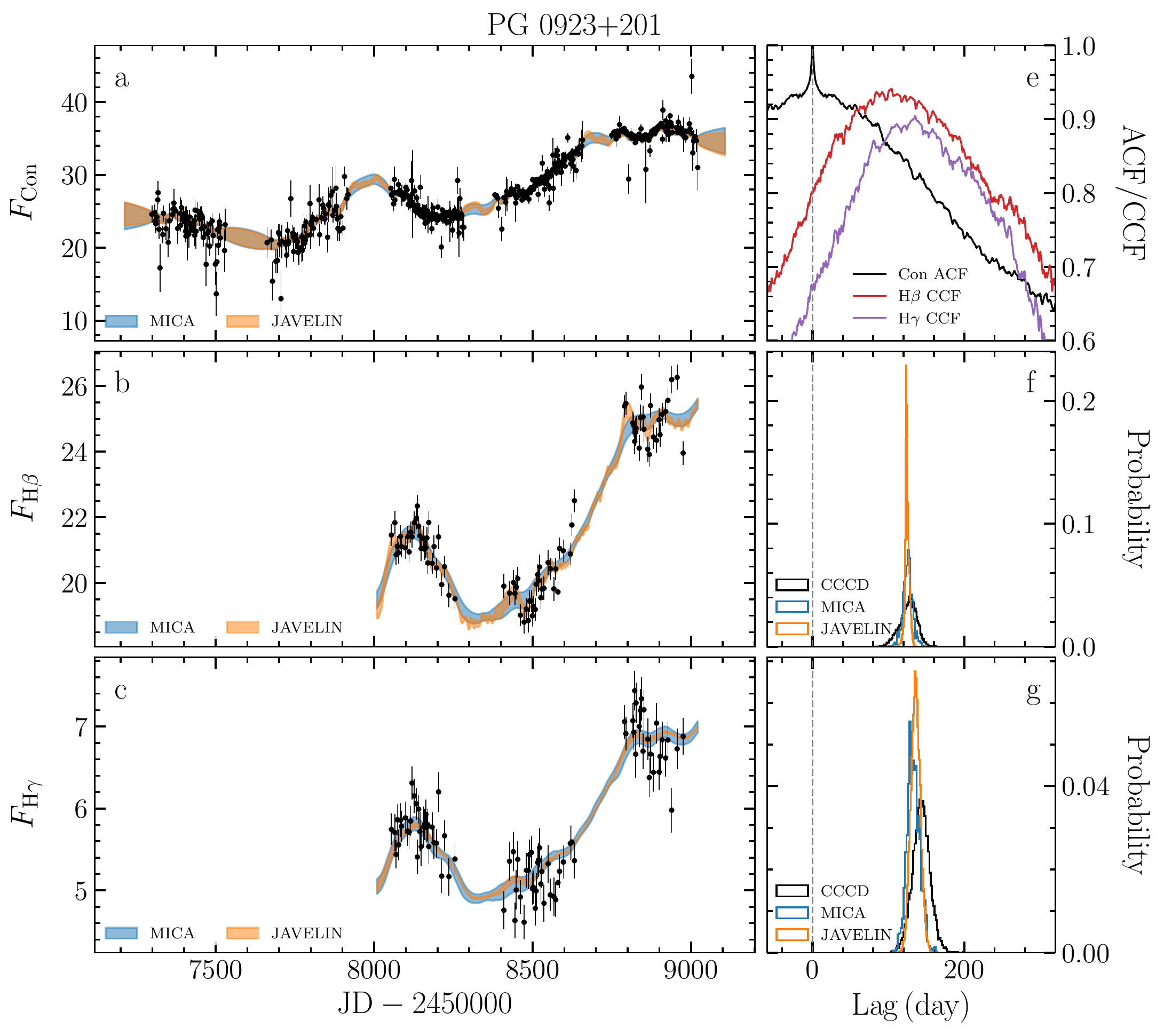}
\caption{Left panels show the light curves of the combined continuum (a) and broad emission lines (b)-(c).
Superposed are the reconstructions by JAVELIN (in orange) and MICA (in blue). The units of the continuum and 
emission line light curves are $10^{-16}~\ergscm$~\AA$^{-1}$ and $10^{-14}~\ergscm$, respectively. Right panels 
show time lag analysis. Panel (e) plots the ACF of the combined continuum, and the CCFs between the broad 
emission lines with continuum light curves. Panels (f)-(g) show the distributions of time delays for the broad 
emission lines.}
\label{fig:lc}
\end{figure*}

\begin{figure*}
\centering
\includegraphics[width=0.8\textwidth]{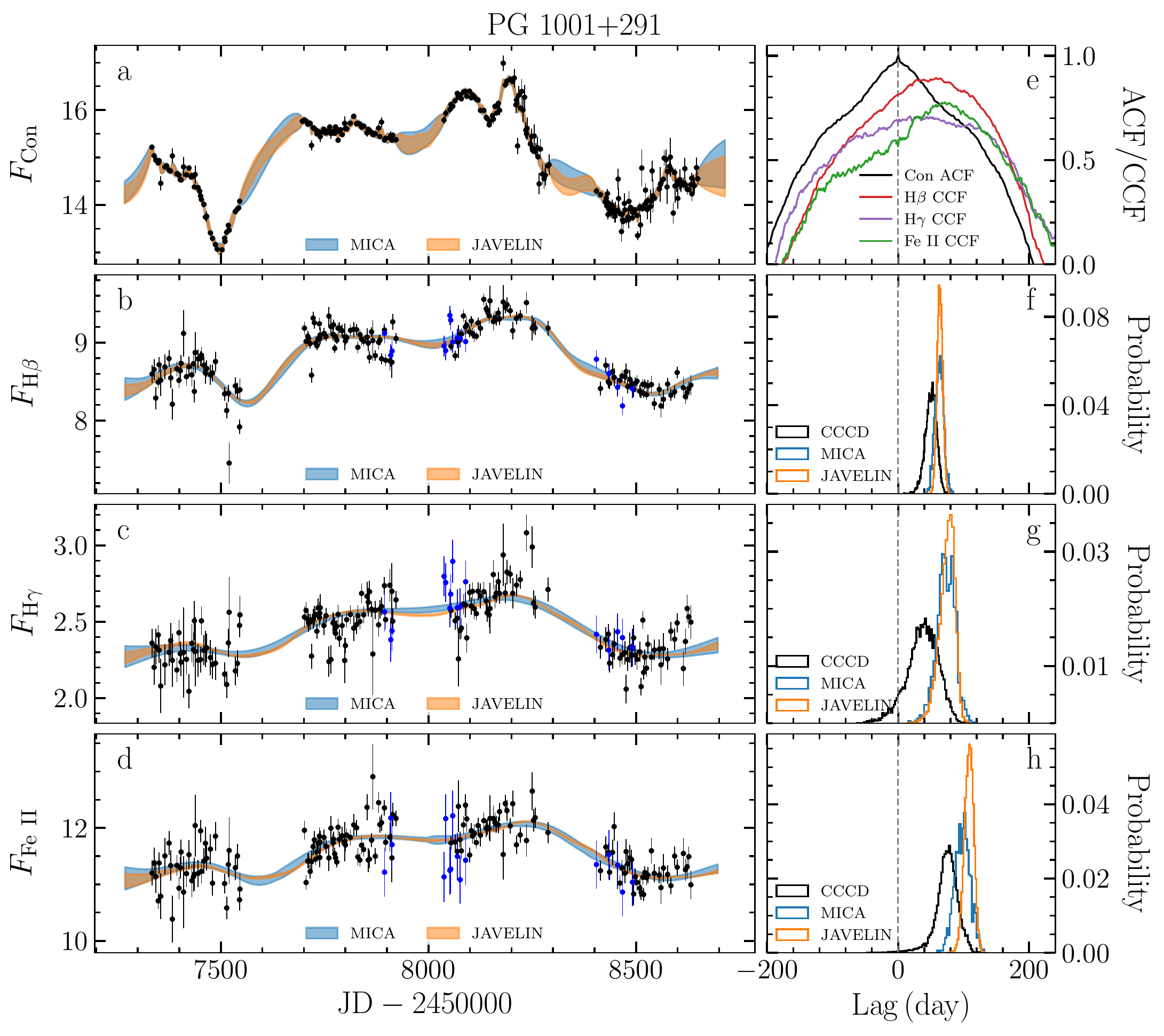}
\caption{Left panels show the light curves of the combined continuum (a) and broad emission lines (b)-(d). 
The black and blue points with error bars are from Lijiang and CAHA observations, respectively. Superposed are 
the reconstructions by JAVELIN (in orange) and MICA (in blue). The units of the continuum and emission line light 
curves are $10^{-16}~\ergscm$~\AA$^{-1}$ and $10^{-14}~\ergscm$, respectively. Right panels show time lag 
analysis. Panel (e) plots the ACF of the combined continuum, and the CCFs between the broad emission lines with 
continuum light curves. Panels (f)-(h) show the distributions of time delays for the broad emission lines.}
\label{fig:lc2}
\end{figure*}

We merge the photometric data from our observations, the ASAS-SN, and the ZTF into the 5100~\AA\, continuum 
to obtain a combined continuum light curve. Time lags between different bands are typically of days (e.g.,
\citealt{Edelson2019}), far smaller than the time lags of broad emissions, therefore, such an intercalibration 
is feasible and does not affect the final time lag measurements.
Due to inhomogeneous apertures, the photometric and 5100~\AA\, continuum data need intercalibration, namely, 
applying additive and multiplicative factors to bring different light curves into a common scale. We use the 
Python package PyCALI \footnote{PyCALI is available at {\url{ https://github.com/LiyrAstroph/PyCALI}.}}, which 
employs a Bayesian framework to do intercalibration \cite[see details in][]{li2014L}. The intercalibrated 
continuum light curves are shown in Figure~\ref{fig:intercali}. For PG~1001+291, the ASAS-SN data are not used 
(see Section~\ref{sec:phot}). After intercalibration, we further rebin the continuum light curves by one day 
apart to combine measurements on the same night. The finally rebinned continuum light curves are used to measure 
time lags, shown in the panel (a) of Figures~\ref{fig:lc} and \ref{fig:lc2}.

For PG~1001+291, we obtained spectra from both Lijiang and CAHA observations, therefore, an intercalibration 
of the spectra is also required. Since the instrumental broadening width (in terms of FWHM) of Lijiang and CAHA 
observations are roughly similar and far smaller than broad line widths, there is no need to adjust the spectral 
resolutions. We simply use the intercalibration factors obtained above to align fluxes of the spectra. In the 
panels (b)-(d) of Figure~\ref{fig:lc2}, we show the intercalibrated light curves of emission lines for 
PG~1001+291 with Lijiang data in black and CAHA data in blue.

\subsection{Variability Characteristics}
As usual, we calculate quantities $F_{\rm var}$ and $R_{\rm max}$ to measure the intrinsic variability 
amplitudes (\citealt{Rodr1997}). Here, $R_{\rm max}$ is the ratio of maximum to minimum fluxes of the light
curve, and $F_{\rm var}$ is defined as 
\begin{equation}
F_{\rm var}=\frac{(S^2-\triangle^2)^{1/2}}{\bar{F}},
\end{equation}
where $S^2$ is the sample variance, $\triangle^2$ is the mean square error, and $\bar{F}$ is the sample mean 
flux
\begin{equation}
S^2=\sum_{i=1}^{N} \frac{(F_i-\bar{F})^2} {(N-1)}, \quad \triangle^2=\sum_{i=1}^{N} \frac{\Delta_i^2}{N},\quad
\bar{F}=\sum_{i=1}^N\frac{F_i}{N},
\end{equation}
where $F_i$ is the flux of $i$th observation,   $\Delta_i$ is the uncertainty of $F_i$, and $N$ is the 
total number of epochs. The standard deviation of $F_{\rm var}$ is estimated by (\citealt{Edelson2002})
\begin{equation}
\sigma_{\rm var}=\frac{1}{F_{\rm var}} \left(\frac{1}{2 \,N} \right)^{1/2}\frac{S^2}{{\bar{F}}^2}.
\end{equation}
The variability characteristics of our light curves are given in Table~\ref{table:stas}. The variability 
amplitudes of the emission lines for both PG~0923+201 and PG~1001+291 can be ranked in the following order: 
H$\gamma$ \textgreater H$\beta$, H$\gamma$ \textgreater H$\beta$ \textgreater \feii. 
Such variability ranks are commonly seen in previous observations (e.g., \citealt{Bentz2010, Barth2015, 
hu2020a, hu2020b}). The mean fluxes of combined continuum and broad emission lines are also listed in 
Table~\ref{table:stas}.

\begin{deluxetable*}{lccccccccrr}
\renewcommand\arraystretch{1.2}
\tabletypesize{\footnotesize}
	\tablecaption{Light curve statistics \label{table:stas}}
	\tablecolumns{5}
	\tablewidth{\textwidth}
		\tablehead{
		\multicolumn{1}{l}{Object} &
			\colhead{Light Curve} &
			\colhead{Mean Flux} &
			\colhead{$F_{\rm var}(\%)$} &
			\colhead{$R_{\rm max}$} &
			\colhead{$\taud$}&
			\colhead{$\sigmad$}
		}
		\startdata
PG~0923+201  & $\rm Con$ & $28.06\pm5.12$ & $17.81\pm0.59$ &3.20&$965\pm303$ & $3.47\pm$0.56\\
             & ${\rm H\beta}$ & $21.78\pm2.14$ & $9.74\pm0.75$ &1.40&$315\pm169$ & $1.88\pm$0.47\\
             & ${\rm H\gamma}$ & $5.84\pm0.73$ & $12.06\pm0.99$ &1.61&$277\pm167$ & $0.62\pm$0.16\\
              \hline
PG~1001+291  & $\rm Con$ & $14.90\pm0.90$ & $5.97\pm0.27$ &1.30 &$279\pm160$ & $0.72\pm$0.19\\
             & ${\rm H\beta}$ & $8.85\pm0.37$ & $4.06\pm0.23$ &1.28 &$135\pm 92$ & $0.35\pm$0.10\\
             & ${\rm H\gamma}$ & $2.46\pm0.19$ & $6.72\pm0.48$ &1.51&$158\pm107$ & $0.15\pm$0.04\\
              & ${\rm Fe\ II}$ & $11.58\pm0.44$ & $3.17\pm0.24$ &1.24&$57\pm38$ & $0.37\pm$0.08\\
		\enddata
		\tablecomments{\footnotesize The continuum fluxes are in a unit of $10^{-16}~\ergscm$~\AA$^{-1}$ 
and the fluxes of all emission lines are in a unit of $10^{-14}~\ergscm$. Fluxes are corrected with the 
Galactic extinction. $ \taud $ is the rest-frame damping timescale in a unit of day and $ \sigmad $ is the
variation amplitude, which has the same unit as the continuum/emission line fluxes (see Section~\ref{sec:simu}).
}
\end{deluxetable*}

\section{Analysis and Results} \label{sec:result}
\subsection{Time Lags}\label{sec:lag}
We calculate time lags between the combined continuum and broad emission-lines flux variations through three 
methods: the interpolated cross-correlation function \citep[ICCF; ][]{Gaskell1986,Gaskell1987}, JAVELIN 
\citep{zu11}, and MICA\footnote{MICA is available at {\url{ https://github.com/LiyrAstroph/MICA2}.}} 
\citep{li2016}. For the ICCF method, the time lag is estimated by $\tau_{\rm cent}$, defined as the centroid 
of the ICCF above 80$\%$ of the peak value ($r_{\rm max}$; \citealt{Peterson2004}). The uncertainties of time
lags are obtained by the $15.87\%$ and $84.13\%$ quantiles of the cross-correlation centroid distributions 
(CCCDs), generated by the ``flux randomization/random subset sampling (FR/RSS)'' method \citep{Peterson1998}. 

Both JAVELIN and MICA use the damped random walk (DRW) model (e.g, \citealt{Kelly2009}) to describe the continuum 
variability and a specific transfer function to fit the light curves of emission lines. JAVELIN adopts a top-hat 
function to approximate the realistic transfer function, while MICA adopts a family of displaced Gaussians. For 
simplicity, we use only one Gaussian in MICA. We assign time lags as the centers of the top-hat/
Gaussian. In MICA, we additionally switch on the functionality of including a parameter for any unknown 
systematic errors, which is added to the data errors in quadrature. 
JAVELIN employs the affine invariant sampling algorithm (\citealt{Goodman2010}; implemented by the package 
\texttt{emcee}\footnote{Available at \url{https://github.com/dfm/emcee}.}) and MICA employs the diffusive 
nested sampling algorithm (\citealt{Brewer2011}; implemented by the package \texttt{cdnest}\footnote{Available 
at \url{https://gitub.com/LiyrAstroph/CDNest}.}) to perform the Markov Chain Monte Carlo (MCMC) technique. 
This generates posterior samples of the model parameters. The time lags and their associated errors are 
estimated from the median, $15.87\%$ and $84.13\%$ quantiles of the corresponding posterior distributions. 
Below, we denote time lags obtained by JAVELIN and MICA as $\tau_{\rm JAV}$ and $\tau_{\rm MICA}$, respectively. 

In Figures \ref{fig:lc} and \ref{fig:lc2}, the left panels show the reconstructed light curves by JAVELIN 
(in orange) and MICA (in blue). The right panels show the autocorrelation functions (ACFs) of the combined 
continuum light curves, ICCFs as well as the time lags distributions in the observed frame.

The measured time lags by the three methods are listed in Table~\ref{table:lag}. We can find general 
agreements to within uncertainties among the results of the three methods. In particular for PG~0923+201, 
the time lags of H$\beta$ and H$\gamma$ measured from ICCF, JAVELIN and MICA are well consistent with each 
other. For PG~1001+291, the H$\beta$, H$\gamma$ and \feii\, lags obtained by JAVELIN and MICA are in good
agreement.

It is worth mentioning that PG~1001+291 was previously monitored by \cite{Du2018} between 2015 and 2017, 
who reported an H$\beta$ time lag of $32.2_{-4.2}^{+43.5}$ days and a mean 5100~\AA\, luminosity of 
$(3.31\pm0.08)\times10^{45}~{\rm erg~s^{-1}}$. The large upper error of the reported H$\beta$ time lag 
was caused by the relatively short temporal baseline and low variation amplitudes of the H$\beta$ light 
curve. By comparison, our campaign obtains a H$\beta$ time lag of $37.3_{-6.0}^{+6.9}$ days and a slightly 
higher mean 5100~\AA\, luminosity of $(3.90\pm0.24) \times10^{45}~{\rm erg~s^{-1}}$ (due to variability). 
The time lag uncertainty of our measurement is more symmetric and significantly reduced. 

\begin{figure}
\centerline{
\includegraphics[width=0.6\textwidth]{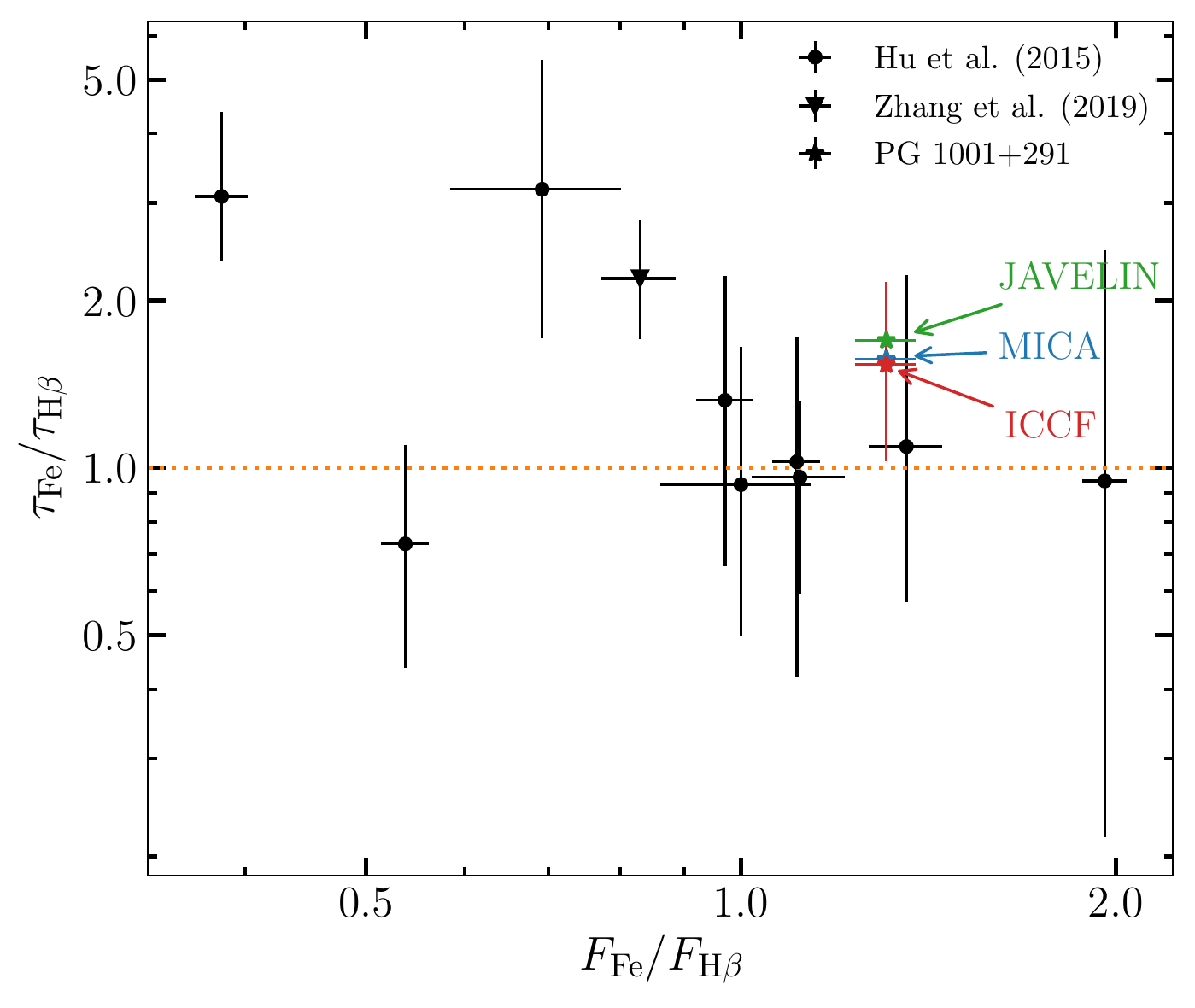}
}
\caption{Relation between time lag ratios of \feii\ to H$\beta$ and intensity ratios of \feii\ to H$\beta$.}
\label{fig:feii-hb}
\end{figure}

Figure~\ref{fig:feii-hb} shows the relationship between time lag ratios of \feii\, to H$\beta$ 
($\rm \tau_{Fe\ II}\, / \tau_{H\beta}$) and flux ratios of \feii\ to H$\beta$ for PG~1001+291 together with 
the samples of \cite{hu2015} and 3C~273 from \cite{Zhang2019}. The results of PG~1001+291 are generally 
consistent with the trend that when $\Rfe>1$, $\rm \tau_{Fe\ II}$ is approximately equal to 
$\rm \tau_{H\beta}$, whereas when $\Rfe<1$, $\rm \tau_{Fe\ II}$ is larger than $\rm \tau_{H\beta}$. 

Through the Monte-Carlo simulation tests described below, we demonstrate that our measured time lags are 
reliable and not caused by seasonal gaps. We adopt the $\tau_{\rm cent}$ values to calculate the black 
hole masses in the following analysis.

\begin{deluxetable*}{lcccccccccc}
\renewcommand\arraystretch{1.4}
\tabletypesize{\scriptsize}
    \tablecaption{Rest-frame time lag and transfer function width measurements\label{table:lag}}
    \tabletypesize{\footnotesize}
    \tablehead{
    \multirow{2}{*}{Object} &
		\multirow{2}{*}{$\rm Line$} &
		\multirow{2}{*}{$r_{\rm max}$} &
		\colhead{} &
		\multicolumn{1}{c}{$\tau_{\rm cent}$} &
		\multicolumn{1}{c}{$\tau_{\rm MICA}$} &
	\multicolumn{1}{c}{$\tau_{\rm JAV}$}&&
	\multicolumn{1}{c}{$W_{\rm MICA}$}&
	\multicolumn{1}{c}{$W_{\rm JAV}$}\\
	\cline{5-7}\cline{9-10}
	\colhead{} &
	\colhead{} &
	\colhead{} &&
	\multicolumn{3}{c}{(day)} &&
	\multicolumn{2}{c}{(day)}
    }
    \startdata
PG~0923+201  & ${\rm H\beta}$ & 0.94&& $108.2_{-12.3}^{+6.6}$ & $105.4 _{-4.9}^{+4.9}$ & $104.3_{-1.4}^{+2.5}$ && $40.2_{-27.9}^{+41.6}~(45.8_{-33.6}^{+35.9})$ &$4.9_{-4.0}^{+15.4}~(11.6_{-10.8}^{+8.6})$\\
 & ${\rm H\gamma}$ & 0.90 && $121.0_{-10.2}^{+8.5}$ & $111.2 _{-6.5}^{+7.5}$  & $113.9 _{-5.3}^{+5.3}$&&$21.8_{-11.7}^{+19.9}~(25.5_{-15.4}^{+16.1})$ & $52.5_{-21.1}^{+15.7}~(50.2_{-18.8}^{+18.0})$\\
\hline  
PG~1001+291 & ${\rm H\beta}$ & 0.89 &&  $37.3_{-6.0}^{+6.9}$ & $47.5_{-5.1}^{+4.7}$ & $48.2_{-3.1}^{+3.3}$&&$68.0_{-9.8}^{+10.4}~(68.3_{-10.1}^{+10.1})$ & $89.1_{-8.0}^{+8.2}~(89.2_{-8.1}^{+8.1})$\\  
& ${\rm H\gamma}$ & 0.71 && $28.0_{-17.6}^{+17.9}$ & $55.1_{-10.3}^{+10.3}$ & $57.4_{-10.2}^{+7.3} $&&$166.9_{-27.0}^{+29.2}~(168.4_{-28.5}^{+27.7})$ &$165.8_{-34.4}^{+26.2}~(163.0_{-31.6}^{+29.0})$\\
 & ${\rm Fe\ II}$ & 0.78 && $57.2_{-11.8}^{+10.5}$ & $74.5_{-9.3}^{+8.4}$ & $81.8_{-5.5}^{+5.0}$ &&$115.7_{-27.7}^{+26.6}~(114.4_{-26.4}^{+27.9})$ &$165.8_{-21.1}^{+19.3}~(164.9_{-20.2}^{+20.2})$\\
\enddata
\tablecomments{``$W_{\rm MICA}$" is the FWHM of the Gaussian transfer function obtained in the MICA fits, and ``$W_{\rm JAV}$" is the width of the top-hat transfer function obtained in the JAVELIN fits. The widths and uncertainties are estimated from the median and 68.3\% confidence levels of the corresponding posterior distributions, respectively. {\cblue{The values in brackets refer to the means with the 68.3\% confidence levels.}}
}
\end{deluxetable*}

\subsection{Validity Tests of ICCF Time Lags}\label{sec:simu}
\begin{figure}
\centering
\includegraphics[width=0.45\textwidth]{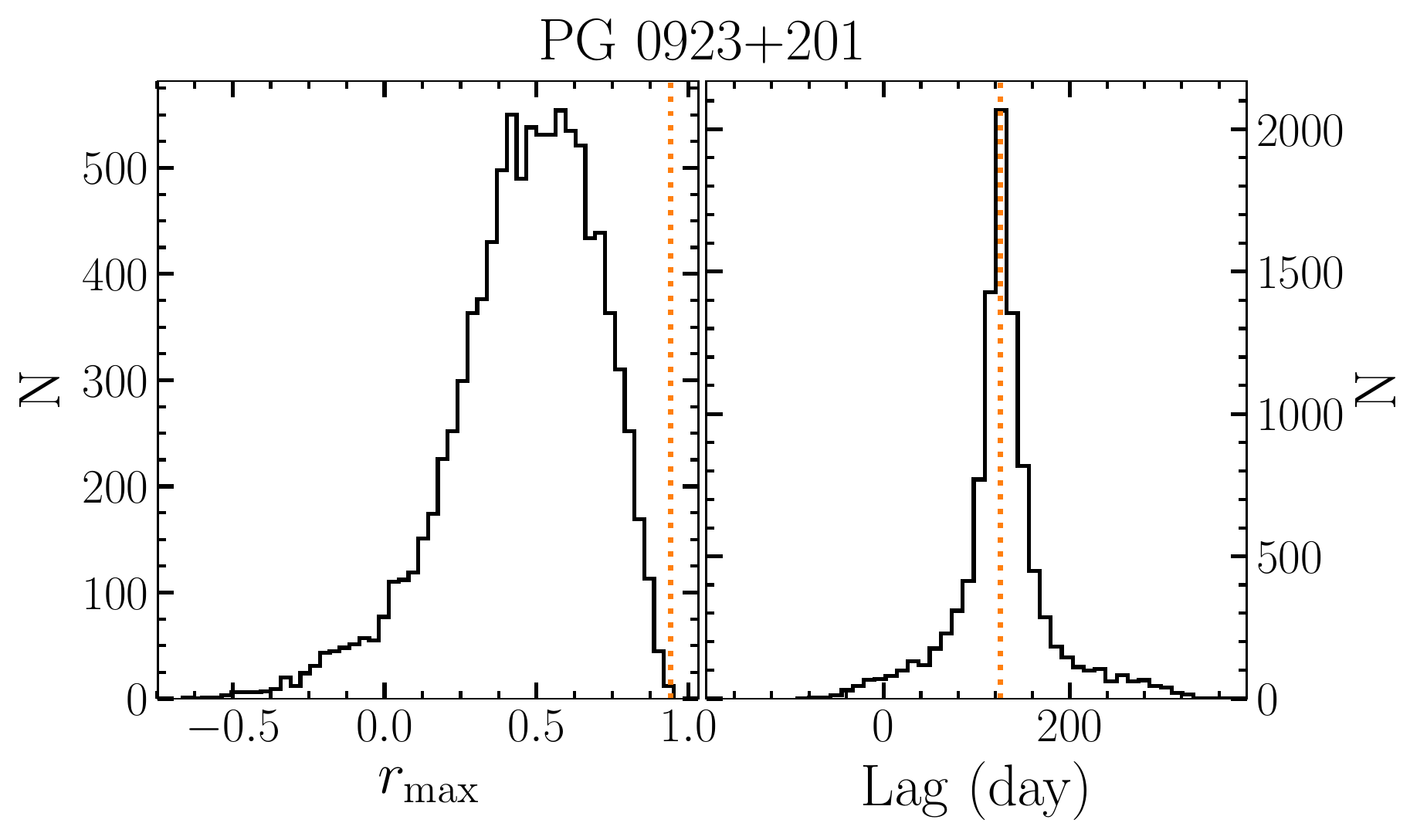}
\includegraphics[width=0.45\textwidth]{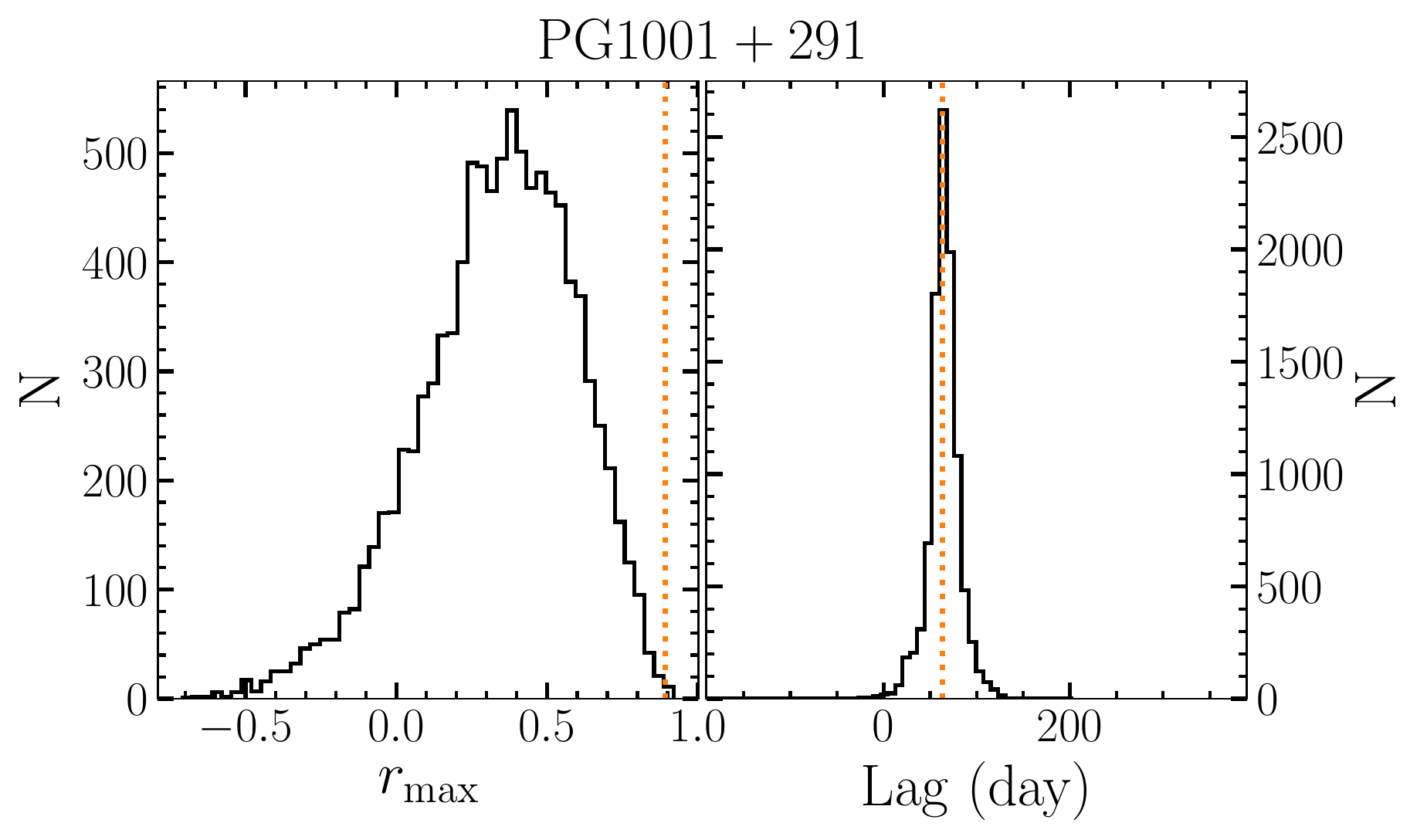}
\caption{Validity tests of H$\beta$ time lags for PG~0923+201 (top) and PG~1001+291 (bottom). Left panels show 
$r_{\rm max}$ distributions of uncorrelated mock light curves from 10000 simulations. The orange dotted lines 
represent $r_{\rm max}$ of the observed light curves. Right panels show  $\tau_{\rm cent}$ distributions of mock 
light curves. The orange dotted lines represent the input time lags (see Section~\ref{sec:simu} for details).}
\label{fig:simu}
\end{figure}

Since our light curves have seasonal gaps, we employ Monte-Carlo simulations to test whether the correlations
between light curves are caused by seasonal gaps. We generate uncorrelated mock light curves based on the 
observed light curves of the continuum and broad emission-line. The mock light curves follow the DRW process, 
which has a covariance between times $t_{i}$ and $t_{j}$ (e.g., \citealt{Kelly2009})
\begin{equation}
S(t_i, t_j) = \sigmad^{2}\, \exp\left( -\frac{|t_{i}-t_{j}|}{\taud}\right),
\end{equation}
where $ \taud $ is a damping timescale and $ \sigmad $ is variation amplitude. We first use the observed light 
curves to determine the parameters $\taud$ and $\sigmad$ and their uncertainties (listed in 
Table~\ref{table:stas}), from which we randomly draw pairs of $\taud$ and $\sigmad$. We then generate two sets 
of mock light curves with daily samplings and use a linear interpolation onto the observed epochs to mimic real
observations. We add Gaussian noises to the mock light curves by enforcing the relative errors equal to these of 
the observed light curves. We finally use the same ICCF method to calculate $r_{\rm max}$ of the two sets of 
light curves. This process is repeated 10000 times and we count the probability that $r_{\rm max}$ is higher than 
that of the observed light curves. We call this the false-alarm probability. For PG~0923+201, the false-alarm 
probabilities of the H$\beta$ and H$\gamma$ are 0.0001 and 0.0005, respectively. For PG~1001+291, the false-alarm 
probabilities of the H$\beta$, H$\gamma$ and Fe~II are 0.0007, 0.0260 and 0.0023, respectively. These quantities 
are relatively low, indicating that the correlations of our light curves are realistic and the influence of 
seasonal gaps is minor. As an example, we show the $r_{\rm max}$ distributions of mock light curves for the 
H$\beta$ in the left panels of Figure~\ref{fig:simu}.

In addition to the above tests, we also design tests to check whether the obtained time lags are reliable. We 
use the same method above to construct a daily sampled continuum light curve, and convolve it with a Gaussian 
transfer function to generate a mock light curve of emission-line. The center and width of the Gaussian transfer 
function (listed in Table~\ref{table:lag}) are randomly assigned according to the best estimates and 
uncertainties obtained by MICA on the observed light curves. The mock light curves are again interpolated onto 
observed epochs. We repeat this process 10000 times and perform the ICCF method to calculate distributions of 
$\tau_{\rm cent}$. Figure~\ref{fig:simu} shows an example of the distributions of H$\beta$ time lags for mock 
light curves, which are well consistent with the input values. The results for the other broad lines are similar, 
implying that our measured time lags are reliable.

\begin{figure*}[th!]
\centering 
\includegraphics[width=0.45\textwidth]{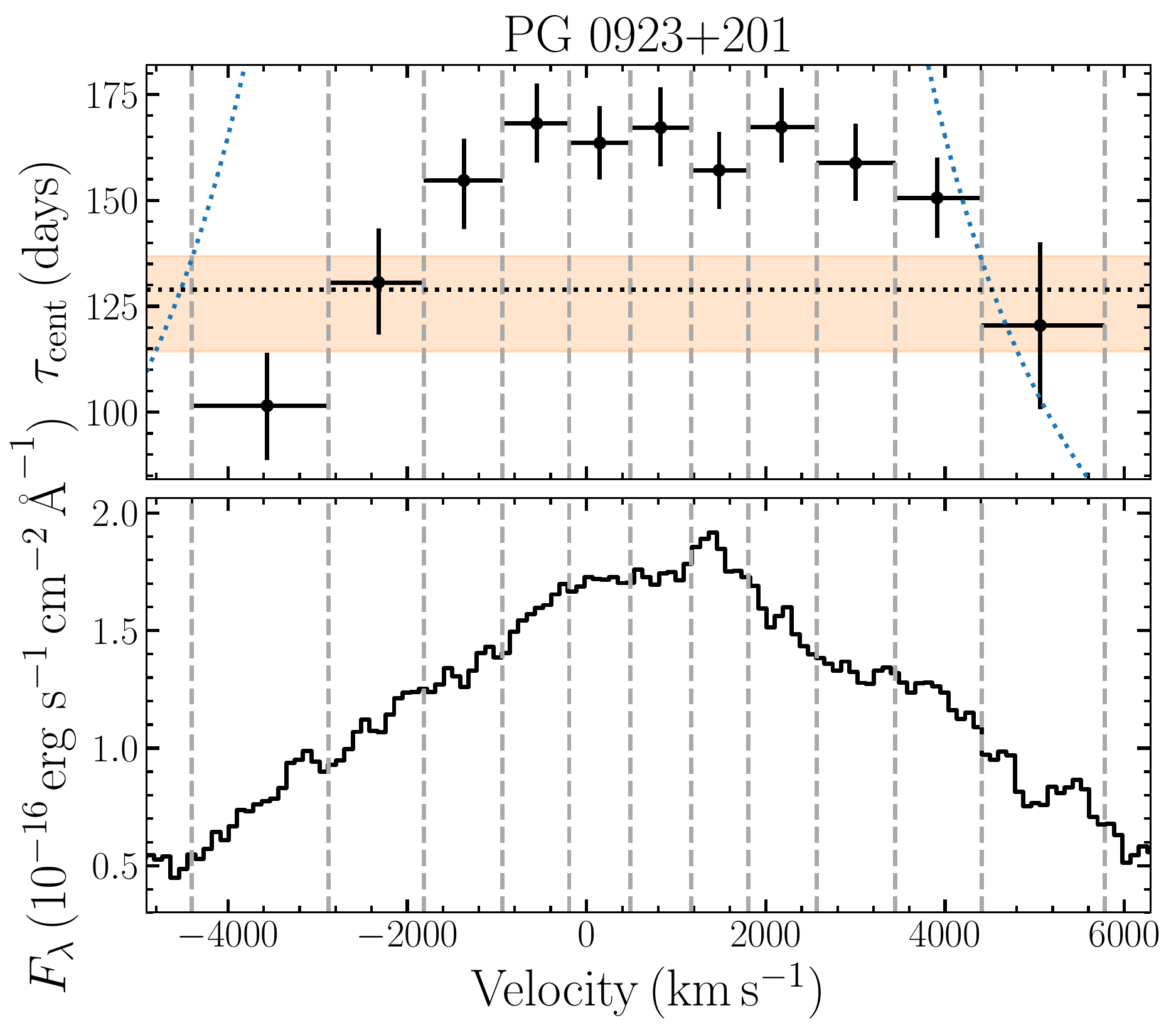}
\includegraphics[width=0.45\textwidth]{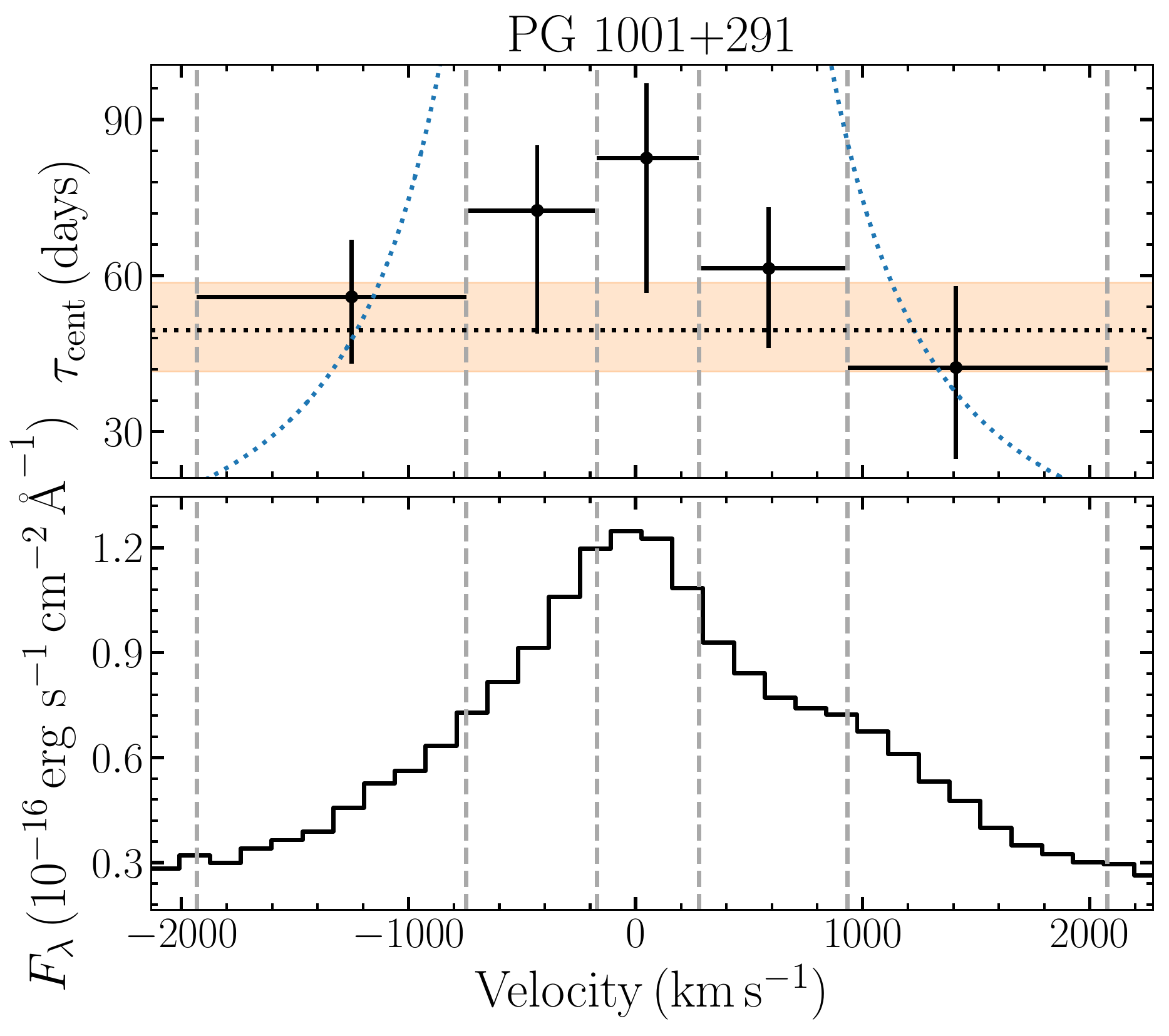}
\caption{Velocity-resolved lags of the H$\beta$ line in PG~0923+201 (left) and PG~1001+291 (right). The top 
panels show delays in the observed frame across velocity bins and the bottom panels show the rms spectra of the 
broad H$\beta$. The vertical dashed lines show the boundaries of velocity bins. The delay in each bin is 
calculated by the ICCF method. The horizontal dotted lines and orange band show the mean H$\beta$ centroid lag 
and the associated uncertainties measured in Section~\ref{sec:lag}. The blue dotted curves show the virial 
envelopes for a Keplerian disk with an inclination of $\cos i = 0.75$ and the estimated black hole mass in 
Table~\ref{tab:fwhm}.} 
\label{fig:BLR}
\end{figure*}
\begin{deluxetable*}{lccccccccccr}
\renewcommand\arraystretch{1.0}
  \tablecolumns{6}
  \setlength{\tabcolsep}{3pt}
  \tablewidth{0.9\textwidth}
  \tablecaption{H$\beta$ line width measurements and black hole mass estimates \label{tab:fwhm}}
  \tabletypesize{\scriptsize}
  \tablehead{
  \colhead{} &
      \multicolumn{2}{c}{Mean Spectrum} &
      \colhead{}    &                    
      \multicolumn{2}{c}{rms Spectrum} & 
      \colhead{}    \\
      \cline{2-3} \cline{5-6}
      \colhead{Object}     &                   
      \colhead{~~~~~~~FWHM~~~~~~~} &                   
      \colhead{~~~~~~~$\sigma_{\rm line}$~~~~~~~}  &
      \colhead{}   &
      \colhead{~~~~~~~FWHM~~~~~~~}       &             
      \colhead{~~~~~~~$\sigma_{\rm line}$~~~~~~~} &
      \colhead{}  &
      \colhead{$\bhm$} &  
      \colhead{$L_{5100}$} &
      \colhead{~~~~$\mathdotM$~~~~} \\
      \colhead{}   &                     
      \multicolumn{2}{c}{(km s$^{-1}$)} &
      \colhead{}  &                      
      \multicolumn{2}{c}{(km s$^{-1}$)} &
      \colhead{} &
      \colhead{~~~~$(10^{7} M_{\odot})$~~~~} &
      \colhead{~~~~$(10^{45} \, \ergs)$~~~~}
      }
\startdata 
PG~0923+201 & $7469\pm 273$ & $3156\pm 109$ & & $6853 \pm 214$ & $2914 \pm 88$ && $118_{-16}^{+11}$& $2.05 \pm 0.29$ & $0.21_{-0.07}^{+0.06}$ \\
PG~1001+291 & $2138\pm 9$ & $1320\pm 4$ & & $2108\pm 120$ & $896 \pm 51$ &&$3.33_{-0.54}^{+0.62}$& $ 3.90 \pm 0.24$ & $679_{-227}^{+259}$\\
\enddata
\end{deluxetable*}

\subsection{Velocity-resolved Delays}
In this section, we calculate velocity-resolved time lags for PG 0923+201 and PG~1001+291, which deliver 
information about the geometry and kinematics of the BLR.

With the spectral decompositions, we extract the broad H$\beta$ component of the model fitted to each individual 
spectrum and calculate the rms spectrum. We then divide the rms spectrum into several velocity bins, each bin 
with the same integrated flux. The light curve of each bin is cross-correlated against the combined continuum to 
obtain time lags. The results are shown in Figure~\ref{fig:BLR}, where the top panels show the velocity-resolved 
delays and the bottom panels show the rms spectra of the broad H$\beta$. In Appendix, we also show the light 
curves and ICCF analysis of each velocity bin. Below we comment on each object individually.

\paragraph{PG~0923+201} The result shows shorter lags at higher velocities on both red and blue wings of the 
line profile, as would be expected for virialized motions. However, the lags in the redshifted bins are overall 
longer than these in the blueshifted bins, which might indicate a signature of outflow \citep{Bentz2009,Du2018a}. 
The virial envelopes in Figure~\ref{fig:BLR} clearly show the asymmetric lag structure. We note that the center 
wavelengths of the narrow lines in our spectra (see Figure~\ref{fig:fit}) are correct, meaning that this 
asymmetry is real. A similar pattern can be seen in other objects, such as NGC~3227 \citep{Denney2009} and 
Mrk~79 \citep{Lu2019}. This complicated structure may imply the coexistence of virilization and outflow in 
the BLR of PG~0923+201.

\paragraph{PG~1001+291} 
The lag structure of PG~1001+291 is relatively symmetric and also shows smaller delays in both wings of the 
line profile, basically agreeing with the virial envelopes. However, this object has a low variability in each 
velocity bin of the H$\beta$ line, resulting in broad ICCFs and large lag errors. Further spectroscopic 
monitoring on PG~1001+291 is warranted to strengthen the evidence on the velocity-delay structure. 

We also use the Monte-Carlo simulation method described in Section~\ref{sec:simu} to calculate the false-alarm 
probability (in terms of $r_{\rm max}$) for each wavelength bin in the delay maps. The resulting false-alarm 
probabilities range between 0.0001 and 0.0197 for PG~0923+201, and between 0.0022 and 0.0116 for PG~1001+291. 
Again, these quantities are relatively low, indicating that the correlations over wavelength bins are not 
dominated by seasonal gaps.

\subsection{Black Hole Masses and Accretion Rates}\label{sec:bhmass}
We use Equation (\ref{equ:bhm}) to calculate the black hole mass. There are two line width measures, namely, 
FWHM and line dispersion ($\sigma_{\rm line}$) of mean or rms spectra. To measure H$\beta$ widths from the mean 
spectrum, we need to subtract the narrow line components. We assume that the narrow H$\beta$ has the same profile 
and a fixed flux ratio of 0.1 to the \OIII\,$\lambda$5007. Regarding uncertainties of the line width, we 
consider the following factors. First, the contribution from the narrow line is estimated by setting a flux ratio 
of 0 and 0.2 between the narrow H$\beta$ to \OIII\,$\lambda$5007. As such, we derive an upper and lower line 
width and assign the uncertainty as the average of the differences to the fiducial value with a flux ratio of 
0.1. Second, we use the bootstrap method to estimate the uncertainties of line widths caused by data sampling
\citep[see, e.g.,][]{Peterson2004}. Specifically, we randomly select $N$ spectra (with replacement) from our $N$ 
total spectra and remove {\cblue{duplicate spectra in creating}} a new mean spectrum. We apply the above 
procedure to this newly generated mean spectrum to calculate its FWHM and $\sigma_{\rm line}$. We repeat this 
process 1000 times to obtain line width distributions, from which we determine the standard deviations and assign 
them as the uncertainties. Finally, for PG~0923+201, we additionally include the influence of the \heii\, by 
assigning the contributed uncertainty as the difference between the line widths with and without adding a \heii\, 
component in the spectral decomposition. We combine the above uncertainties in quadrature to get the final line 
width uncertainties.

For the rms spectrum, we determine line widths by spectral fitting, as shown in the bottom panels of 
Figure~\ref{fig:fit}. The fitting components for the rms spectrum in PG~0923+201 include: (1) a single power 
law, (2) a single Gaussian for H$\beta$, \heii, H$\gamma$; and in PG~1001+291 include: (1) a single power law,
(2) an \feii\ template from \cite{Boroson1992}, (3) a single Gaussian for H$\beta$ and H$\gamma$. The
corresponding uncertainties are determined by the same bootstrap method as applied for the mean spectrum 
described above.

From all measured line widths we subtract in quadrature the instrumental broadening (Section~\ref{sec:spec}) 
to obtain the results in Table~\ref{tab:fwhm}. Following \cite{Du2018}, we adopt $\tau_{\rm cent}$ and the 
H$\beta$ FWHM from the mean spectrum and $\fBLR = 1$ to measure the black hole mass. The mass errors simply
include the uncertainties of time lags and line widths. We obtain a black hole mass of 
$118_{-16}^{+11}\times10^{7} M_{\odot}$ for PG~0923+201 and $3.33_{-0.54}^{+0.62}\times 10^{7} M_{\odot}$ 
for PG~1001+291. 

We also measure the black hole mass using the broad H$\gamma$ line. Because the velocity width and shift of 
H$\gamma$ are fixed to those of H$\beta$ in fitting the mean spectrum, we resort to the FWHM from the rms
spectrum. For PG~0923+201 and PG~1001+291, the FWHMs of H$\gamma$ are $7572\pm 404$ and 
$1884\pm 250$~km~s$^{-1}$, respectively. Again, using the width and $\tau_{\rm cent}$ of H$\gamma$ and 
$\fBLR = 1$, we estimate the black hole mass to be  $135_{-18}^{+17}\times10^{7}M_{\odot}$ for PG~0923+201 and 
$1.94_{-1.32}^{+1.34}\times10^{7} M_{\odot}$ for PG~1001+291 respectively. The obtained black hole masses from 
the H$\beta$ and H$\gamma$ are consistent with each other. 

According to the standard accretion disk model \citep{Shakura1973}, the dimensionless accretion rate 
(Eddington ratio) defined as $\mathdotM =\dot{M}_{\bullet}\,c^2/L_{\rm Edd}$ can be estimated by \citep{Du2016b}
\begin{equation}\label{equ:accre}
\mathdotM = 20.1\,\left(\frac{\ell_{44}}{\cos i}\right)^{3/2}m_7^{-2},
\end{equation}
where $\dot{M}_{\bullet}$ is the accretion rate, $L_{\rm Edd}$ is the Eddington luminosity, 
$\ell_{44}=L_{5100}/10^{44}{\rm \ergs}$, $m_7=\bhm/10^7\sunm$, and $i$ is the inclination angle of the 
accretion disk. We take ${\cos i} = 0.75$, which represents a mean disk inclination for a type 1 AGN by 
assuming that the inclination randomly distributes between 0 and 60 degrees. We estimate the flux contributions 
from the host galaxies based on the empirical relation proposed by \cite{Shen2011}, which is written as 
$L_{\rm 5100,host}/L_{\rm5100,AGN}=0.8052-1.5502x+0.912x^2-0.1577x^3$, for $x<1.053$,  where 
$L_{\rm 5100,host}$ and $L_{\rm5100,AGN}$ are the luminosity of host and AGN at 5100~\AA\,\,respectively, 
$x=\log \left(L_{\rm 5100,tot}/10^{44}{\ergs}\right)$, and $L_{\rm 5100,tot}$ is the total luminosity at
5100~\AA. For $x>1.053$, the luminosity correction is unnecessary. For PG~0923+201 and PG~1001+291, $x=1.380$ 
and $1.623$, respectively, meaning that the contribution of host galaxies can be neglected. This also
demonstrates that it is reasonable to omit the host galaxy component in our spectral decompositions 
in Section~\ref{sec:sfs}. We calculate the mean flux of the continuum light curve obtained by the spectral 
fitting and use it to compute the luminosity at 5100~\AA. Combining the $\bhm$ and $L_{5100}$, we estimate 
$\mathdotM$ to be $0.21_{-0.07}^{+0.06}$ for PG~0923+201 and $679_{-227}^{+259}$ for PG~1001+291.
This implies that PG~0923+201 is a sub-Eddington accretor whereas PG~1001+291 is a super-Eddington accretor. 

In the above calculations, we adopt a virial factor of $\fBLR = 1$. Since the two objects have significantly
different properties, their virial factors might be different. Several previous studies proposed that the 
virial factor $\fBLR$ for H$\beta$ line is anticorrelated with the FWHM (e.g., \citealt{Mej2018,Yu2019,
Aldama2019}). Using the relation of \cite{Mej2018} 
\begin{equation}
f^{c}_{\mathrm{BLR}}=\left(\frac{\mathrm{FWHM}}{4550 \pm 1000}\right)^{-1.17},
\end{equation}
we obtain the virial factor based on the FWHM from the mean spectrum $0.56\pm 0.14$ for PG~0923+201, and 
$2.42 \pm 0.62$ for PG~1001+291. With this new viral factor, the black hole masses are estimated to be 
$66_{-19}^{+18}\times10^{7} M_{\odot}$ for PG~0923+201 and $8.05_{-2.44}^{+2.55}\times10^{7} M_{\odot}$ 
for PG~1001+291. Correspondingly, the dimensionless accretion rates are changed to $0.66_{-0.40}^{+0.38}$ 
and $116_{-71}^{+74}$, respectively. This still indicates that PG~0923+201 is a sub-Eddington accretor 
and PG~1001+291 is a super-Eddington accretor, retaining the conclusion with $\fBLR=1$.

\begin{figure*}[th!]
\centering 
\includegraphics[width=0.45\textwidth]{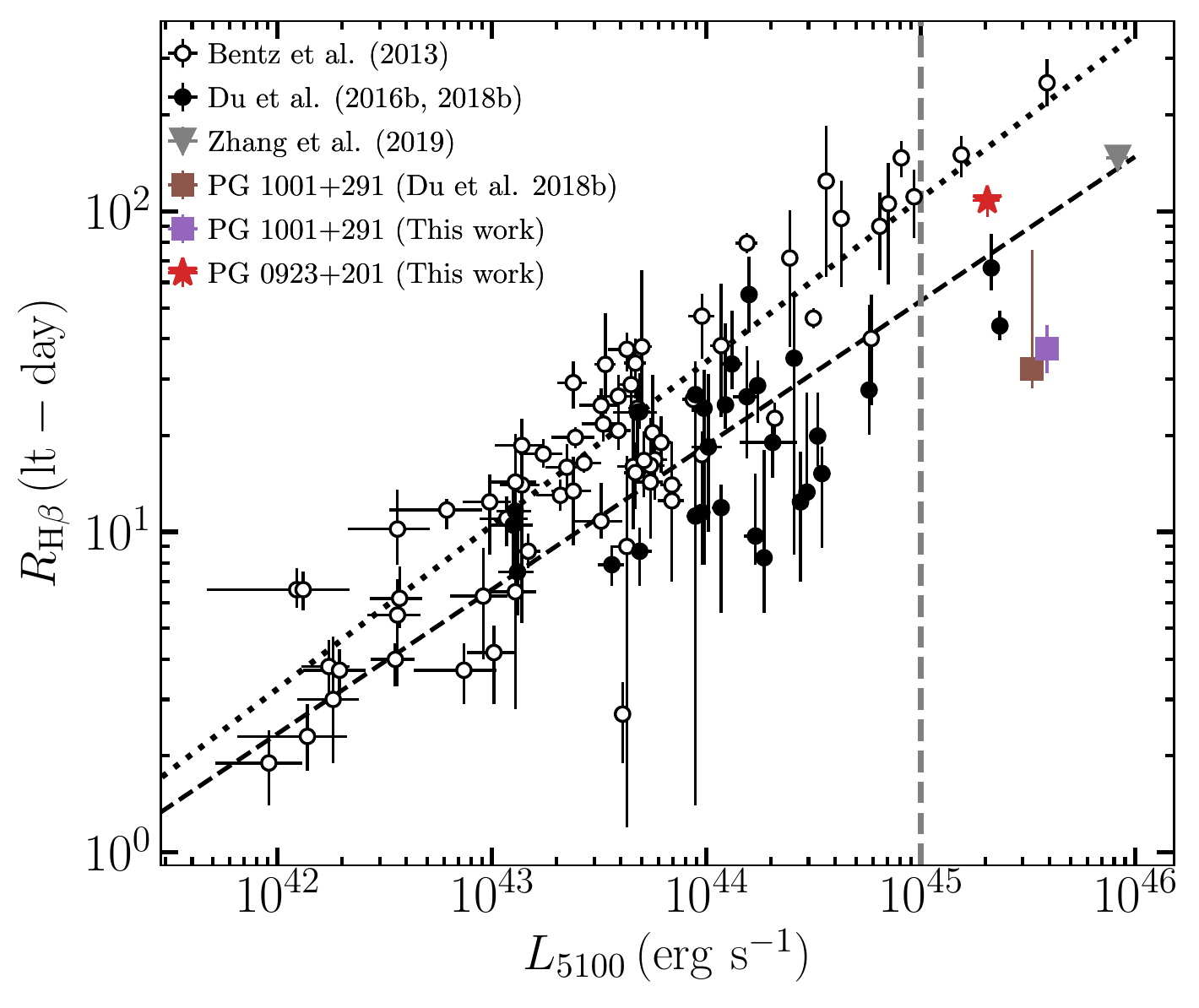}
\includegraphics[width=0.45\textwidth]{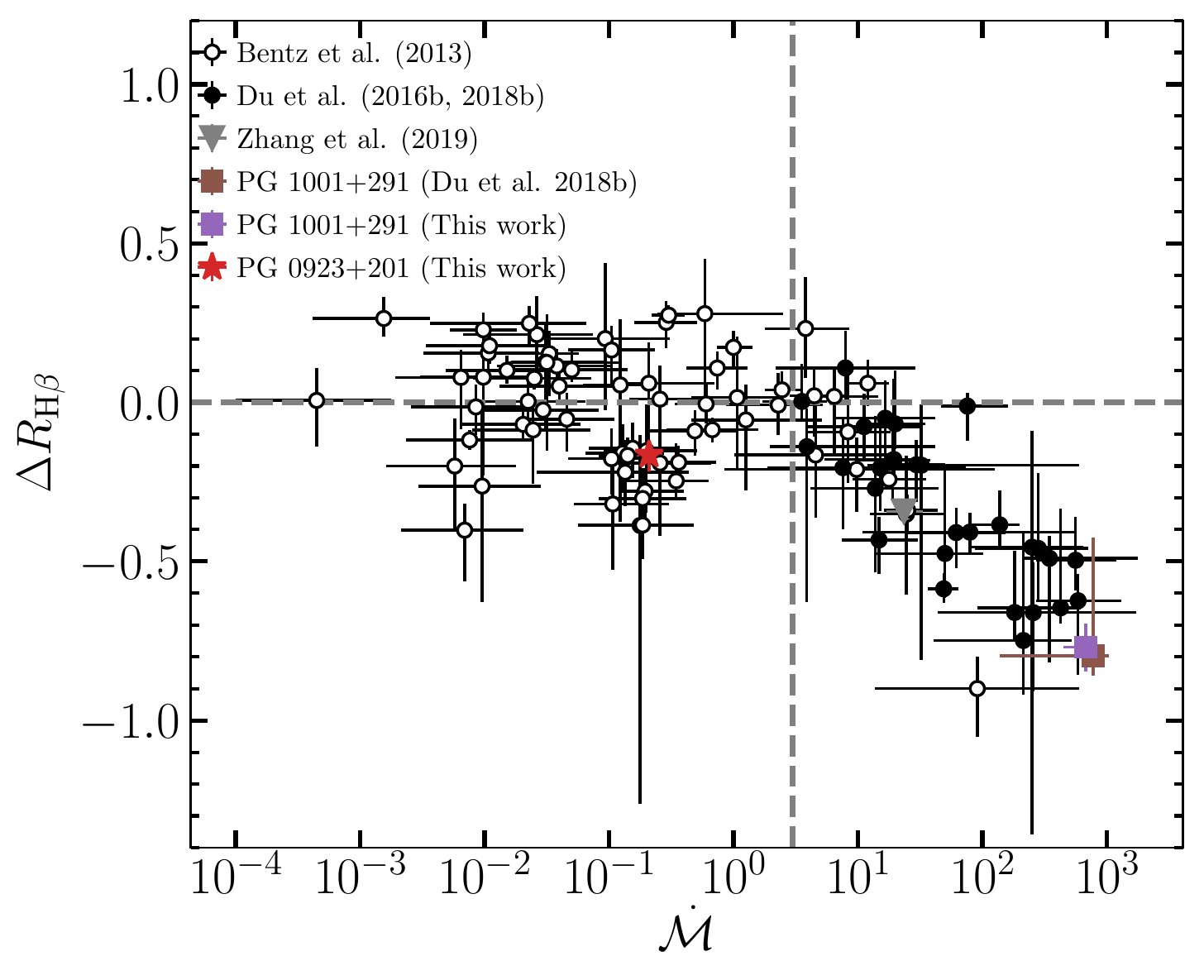}
\caption{(Left) the $\RL$ relation. The dotted and dashed lines are the best fit to the AGNs with 
$\mathdotM<3$ and with $\mathdotM \geq 3$ from \cite{Du2018}. The RM objects with luminosities above 
$10^{45}~{\rm erg~s^{-1}}$ are are located to the right of the vertical dashed line. (Right) the relation 
between $\Delta R_{\rm H\beta}$ and $\mathdotM$. The vertical and horizontal dashed lines indicate 
$\mathdotM=3$ and $\Delta R_{\rm H\beta} = 0$, respectively (see Section \ref{sec:rl} for the definition 
of $\Delta R_{\rm H\beta}$).
}
\label{fig:RL}
\end{figure*}

\section{Discussion}\label{sec:diss}
\subsection{Implication for $\RL$ Relation} \label{sec:rl}
In the left panel of Figure~{\ref{fig:RL}}, we plot the H$\beta$ lag and $L_{5100}$ of PG~0923+201 and 
PG~1001+291 along with the previous compiled sample of \cite{Bentz2013} and samples from the SEAMBH 
campaigns \citep{Du2015,Du2016b,Du2018}. There are seven RM AGNs with luminosities above $10^{\rm 45}~\ergs$ 
to date. For comparison, we also superimpose the empirical $\RL$ relation from \cite{Du2018}. The dotted 
line represents the relation for RM sample with $\mathdotM \!<\! 3$, which is consistent with 
\cite{Bentz2013}'s compilation. The dashed line represents the relation for RM sample with 
$\mathdotM \!\ge\!3$. The location of PG~1001+291 is almost unchanged compared to the previous measurement 
of \cite{Du2018}. The H$\beta$ time delay of PG~0923+201 is consistent with the empirical $\RL$ relations 
of both \cite{Bentz2013} and \cite{Du2018}, in consideration of their associated scatters. However, 
PG~1001+291 lies 0.78~dex below the empirical $\RL$ relation of \cite{Bentz2013}. The reported scatter 
$\sigma$ of the \cite{Bentz2013} relation is about 0.19 and the deviation of PG~1001+291 exceeds a $4\sigma$
significance. This deviation is believed to be caused by the physical dependence of the $\RL$ relation on the 
dimensionless accretion rates. According to the results of \cite{Du2015,Du2016b,Du2018}, there is a strong 
correlation between time lag shortening in SEAMBHs and accretion rates. A possible explanation for time lag 
shortening was proposed by \cite{Wang2014}. Specifically, slim accretion disks with super-Eddington accretion 
rates produce strong self-shadowing effects, resulting in a strongly anisotropic radiation field and two 
dynamically distinct BLR regions with different delays \citep{Wang2014}. The BLR is thereby divided into a 
shadowed region and an unshadowed region \citep{Du2018}. Since the shadowed region receives fewer ionizing 
photons, its size shrinks, leading to a shortened time delay. Another possible explanation {\cblue was that 
the time lag shortening is caused by the changes of the UV/optical spectral energy distribution and the 
relative amount of ionizing radiation, which may be related to the black hole spin and accretion rate 
(e.g.,  \citealt{Wang2014a, Czerny2019, Fonseca2020}).}   

According to the $\RL$ relation of \cite{Du2018},
\begin{equation}  
\log\left(\frac{\RBLR}{{\rm ltd}}\right)\! =\! 
\left\{\! \begin{array}{ll}  
             \!(1.53\pm0.03) \!+\! (0.51\pm0.03)\log \ell_{44}, & \!(\mathdotM \!<\! 3) \\ \\ 
             \!(1.26\pm0.04) \!+\! (0.45\pm0.05)\log \ell_{44}, & \!(\mathdotM \!\ge\! 3)   
             \end{array}\right.
\end{equation}
we can find that PG~0923+201 and PG~1001+291 are located within $2\sigma$ of the relation with 
$\mathdotM \!<\! 3$ and $3\sigma$ of the relation with $\mathdotM \!\ge\! 3$, respectively. This confirms 
that the $\RL$ relation at the high-luminosity end still strongly depends on the dimensionless accretion 
rates. Following \cite{Du2015}, we define $\Delta R_{\rm H\beta} = \rm log(R_{\rm H\beta}/R_{\rm {H\beta},R-L})$ 
to measure the deviation from the $\RL$ relation for $\mathdotM<3$ and plot $\Delta R_{\rm H\beta}$ as a function 
of the dimensionless accretion rate $\mathdotM$ in the right panel of Figure~\ref{fig:RL}. PG~1001+291 clearly 
deviates from the canonical relation by about 0.8 dex. Figure~\ref{fig:RL} demonstrates that H$\beta$ time lags 
are more severely shortened for higher accretion rates. Therefore, it is important to take accretion rates into 
account when estimating the black hole mass of luminous AGNs at super-Eddington accretion rates.

\subsection{Locations in the Eigenvector 1 Plane of RM Samples}
\begin{figure}
\centering 
\includegraphics[width=0.6\textwidth]{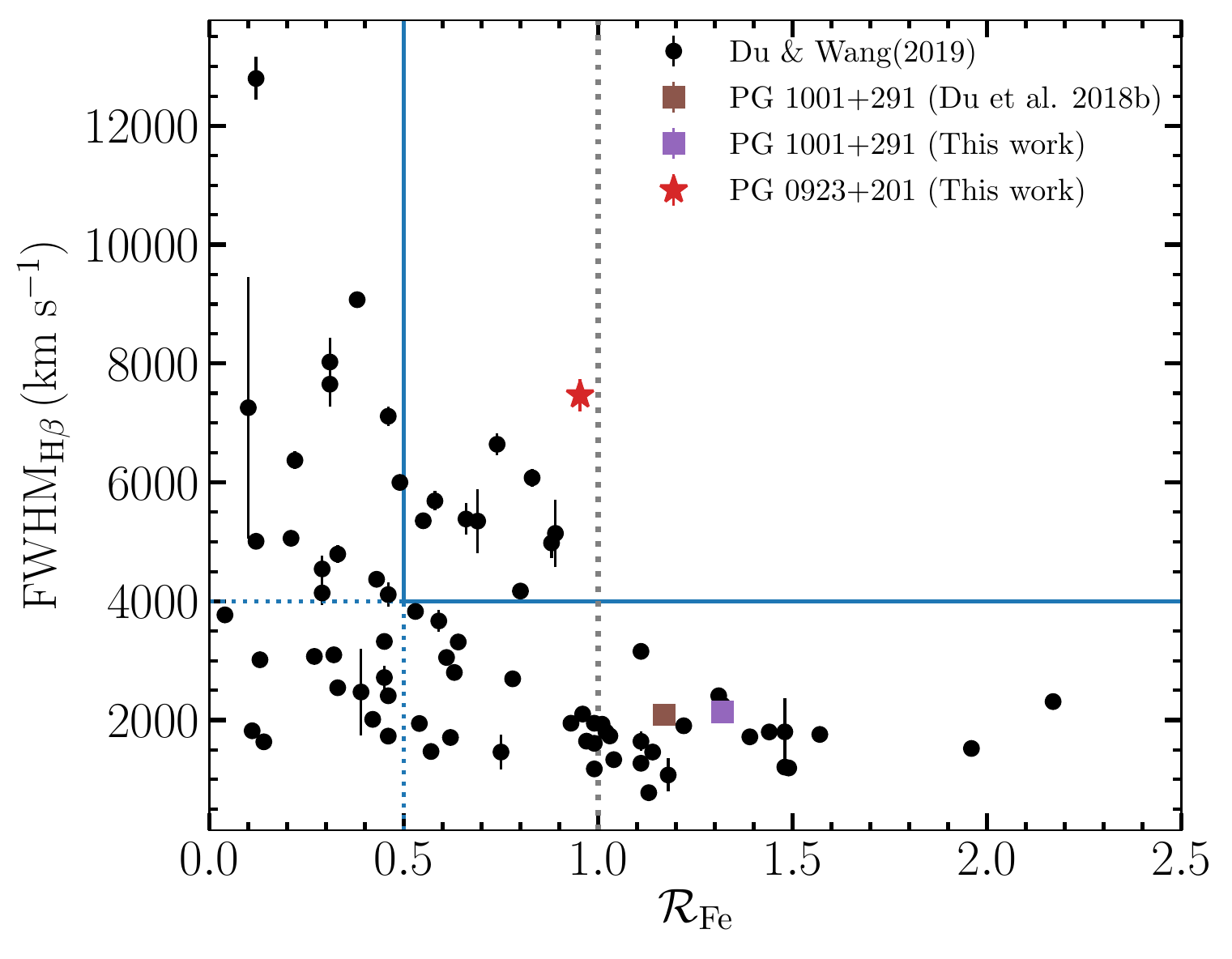}
\caption{The main sequence of RM samples. The black points were compiled by \cite{Du2019}. The blue 
horizontal and vertical lines corresponds to $\rm FWHM_{H\beta} = 4000~km~s^{-1}$ and $\Rfe=0.5$, 
respectively, and the gray dotted line represent $\Rfe = 1.0$. The upper right corner enclosed by the 
blue solid lines is defined as the ``outlier'' region according to \cite{Sulentic2011}.}
\label{fig:Rfe_fwhm}
\end{figure}

In Figure~\ref{fig:Rfe_fwhm}, we plot the distribution of RM samples in the Eigenvector 1 (EV1) plane (also 
known as the main sequence) by including the compilation of \cite{Du2019} and two objects obtained in this 
work. Here, the EV1 plane refers to the \Rfe\, versus $\rm FWHM_{H\beta}$ (FWHM of the broad H$\beta$) plane.
\cite{Sulentic2000} divides AGNs into populations A and B according to the value of $\rm FWHM_{H\beta}$, where 
AGNs with $\rm FWHM_{H\beta} \leq 4000~km~s^{-1}$ are classified as Population A, otherwise as Population B. 
Population A includes NLS1s and high accretors \citep{Marziani2014}, while Population B has larger black hole 
masses and lower accretion rates \citep{Sulentic2011}. It is generally believed that the accretion rate and 
Eddington ratio increase with \Rfe, and the dispersion of $\rm FWHM_{H\beta}$ for a given \Rfe\, characterizes 
the orientation effect \citep{Marziani2001, Shen2014}. PG~1001+291 is located in the region of high \Rfe\, and 
narrow $\rm FWHM_{H\beta}$, consistent with our results that PG~1001+291 is accreting at a super-Eddington rate. 
We note that the \Rfe\, of PG~1001+291 is larger than the previous measurement of \cite{Du2019} due to 
variability. 

We simply follow \cite{Sulentic2011} to define outliers in the Eigenvector 1 plane, namely, those AGNs
with $\rm FWHM_{H\beta} > 4000~km~s^{-1}$ and $\Rfe\,>0.5$. PG~0923+201 is the most significant outlier in 
the present RM sample. The reasons for such {\cblue an ``outlier'' location of PG 0923+201} in the EV1 plane 
are not yet clear. However, we note that this definition of ``outlier'' is only phenomenological. 
Meanwhile, we cannot exclude the possibility that the significant deviation might be caused by real differences 
between PG~0923+201 and the RM sample of \cite{Du2019}. The majority of the \cite{Du2019} sample have low 
luminosities and black hole masses and therefore the corresponding FWHMs are generally small. Nevertheless, 
from the large SDSS quasar sample, there also exists a population of AGNs that have relatively large \Rfe\ 
and broad H$\beta$ line widths like PG~0923+201 (see \citealt{Shen2014}). Detailed multiwavelength 
investigations of these quasars might help reveal additional BLR and SMBH accretion physics.

\section{Conclusion}\label{sec:summary}
We present a reverberation mapping campaign of two luminous quasars at the high-luminosity end of the 
$\RL$ relation. Our main results are as follows.

\begin{itemize}

\item We measure the time delays of the broad emission lines with respect to the 5100~\AA\, continuum. 
Using the ICCF method, the H$\beta$ and H$\gamma$ lags of PG~0923+201 in the rest frame are 
$108.2_{-12.3}^{+6.6}$ and $121.0_{-10.2}^{+8.5}$~days, respectively, and the H$\beta$, H$\gamma$, \feii\ 
lags of PG~1001+291 are $37.3_{-6.0}^{+6.9}$, $28.0_{-17.6}^{+17.9}$ and $57.2_{-11.8}^{+10.5}$~days, 
respectively.

\item The velocity-resolved delays of the H$\beta$ line in PG~0923+201 show a virialized motion, with 
shorter lags at line wings and longer lags at line core. However, the lags in the redshifted bins are 
higher than those in the blueshifted bins. This complicated structure may indicate the coexistence of 
virialized motion and outflow in the BLR of PG~0923+201. The lag structure of PG~1001+291 is relatively 
symmetric and also shows a virialized BLR.

\item Based on the H$\beta$ delays and FWHMs in the mean spectra, and assuming a virial factor of 
$f_{\rm BLR}=1$,  we estimate the black hole masses to be $118_{-16}^{+11}\times10^{7} M_{\odot}$ for 
PG~0923+201 and  $3.33_{-0.54}^{+0.62}\times10^7 M_{\odot}$ for PG~1001+291. We obtain consistent mass 
estimates using the H$\gamma$ line. The accretion rates of PG~0923+201 and PG~1001+291 are estimated to 
be $0.21_{-0.07}^{+0.06}\,L_{\rm Edd}\,c^{-2}$ and $679_{-227}^{+259}\,L_{\rm Edd}\,c^{-2}$, respectively, 
indicating that PG~0923+201 is accreting at a sub-Eddington rate whereas PG~1001+291 is a super-Eddington 
accretor.

\item The H$\beta$ time lag of PG~1001+291 falls 0.78~dex below the empirical $\RL$ relation 
of \cite{Bentz2013}, confirming that even for high-luminosity quasars, H$\beta$ time lags depend on both
luminosity and Eddington ratio, as previously found by \cite{Du2018}. This strengthens the conclusion that 
the $\RL$ relation at the high-luminosity end needs to consider the influences of accretion rates. The
uncomfortably high single-epoch black hole masses estimated for AGN at large redshifts may be significantly 
over-estimated if this effect has not been taken into account.
\end{itemize}

{\cblue We thank the referee for useful comments that improved the manuscript.} This work is supported by 
the National Key R\&D Program of China (2016YFA0400701, 2016YFA0400702), by the National Science Foundation 
of China (NSFC-11721303, 11773029, 11833008, 11873048, 11922304, 11973029, 11991051, 11991052, 11991053, 
11991054, 12003036, 12022301), by the Key Research Program of Frontier Sciences of the Chinese Academy of 
Sciences (CAS; YZDJ-SSW-SLH007), by the CAS Key Research Program (KJZDEW-M06), by the CAS  International 
Partnership Program (113111KYSB20200014), and by the Strategic Priority Research Program of the CAS (XDB23000000, 
XDB23010400). We acknowledge the support of the staff of the Lijiang 2.4 m telescope. Funding for the telescope 
has been provided by the CAS and the People’s Government of Yunnan Province. We also acknowledge the support of 
the staff of the CAHA 2.2 m telescope. Y.-R.L. acknowledges financial support from the Youth Innovation 
Promotion Association CAS. K.H. acknowledges support from STFC grant ST/R000824/1.
Based on observations obtained with the Samuel Oschin Telescope 48-inch and the 60-inch Telescope at the 
Palomar Observatory as part of the Zwicky Transient Facility project. ZTF is supported by the National Science 
Foundation and a collaboration including Caltech, IPAC, the Weizmann Institute for Science, the Oskar Klein 
Center at Stockholm University, the University of Maryland, Deutsches Elektronen-Synchrotron and Humboldt 
University, Lawrence Livermore National Laboratory, the TANGO Consortium of Taiwan, the University of Wisconsin 
at Milwaukee, Trinity College Dublin, and Institut national de physique nucl\'eaire et de physique des 
particules. Operations are conducted by COO, IPAC and University of Washington. This research has made use of 
the NASA/IPAC Extragalactic Database (NED) which is operated by the Jet Propulsion Laboratory, California 
Institute of Technology, under contract with the National Aeronautics and Space Administration.

\textit{Software:} DASpec (\url{ https://github.com/PuDu-Astro/DASpec}), PyCALI \citep{li2014L}, 
MICA \citep{li2016}, JAVELIN \citep{zu11}.

\appendix
\section{Spectral Decompositions with Including the \heii\, Component} \label{sec:add-heii-lc}
\begin{figure}[th!]
\centering 
\includegraphics[width=0.6\textwidth]{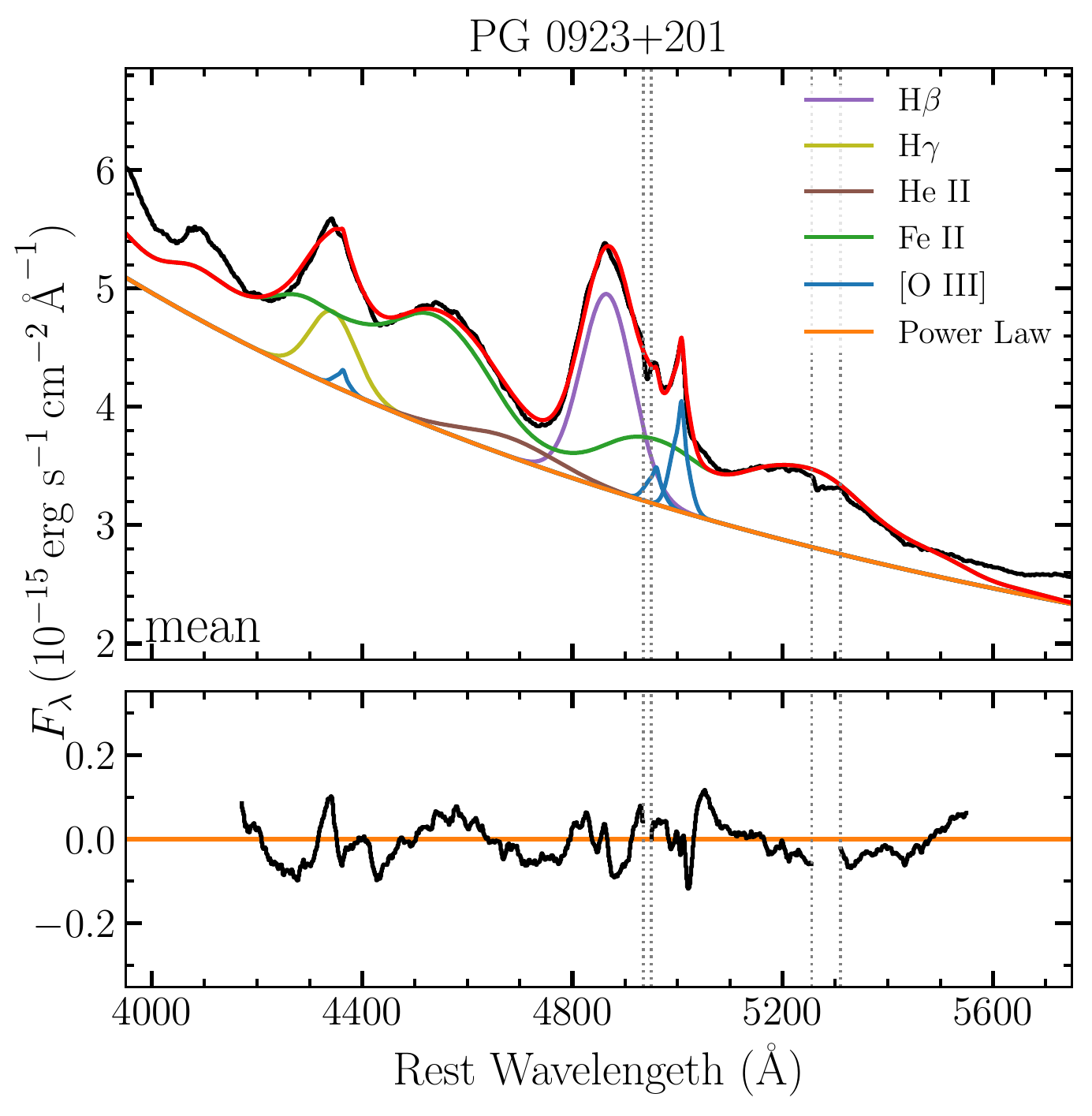}
\caption{\cblue A spectral decomposition of the mean spectrum for PG~0923+201 with including a \ion{He}{2} 
$\lambda4686$ component. The top and bottom panels show the details of the spectral decomposition and the
residuals in the fitting window, respectively. The vertical dotted lines indicate the regions omitted from 
the fit.}
\label{fig:fit-heii}
\end{figure}

\begin{figure}[th!]
\centering
\includegraphics[width=0.8\textwidth]{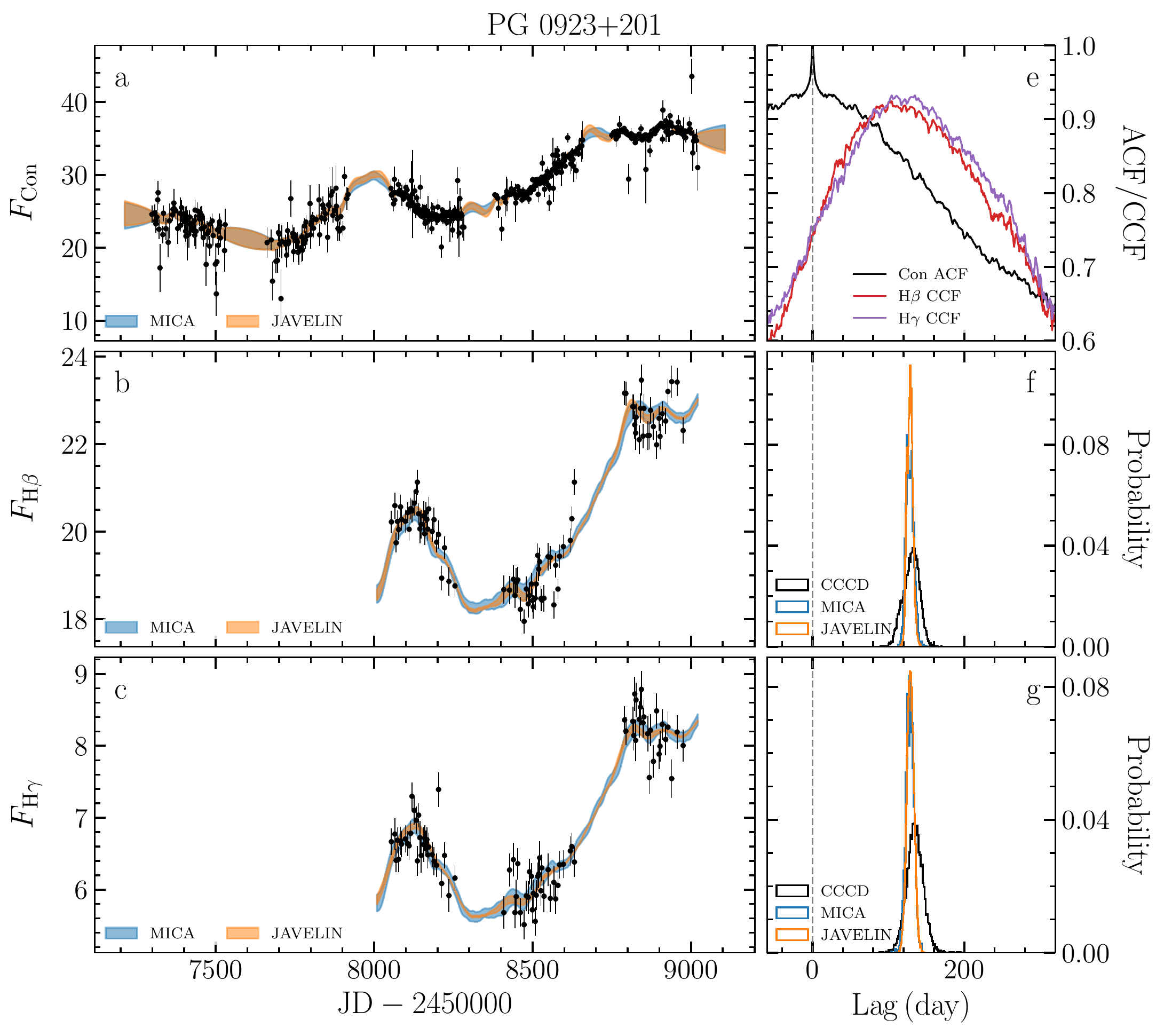}
\caption{\cblue Same as Figure~\ref{fig:lc}, but for spectral decompositions including a \heii\, component.}
\label{fig:lc-heii}
\end{figure}

In this appendix, we present the fitting results of PG~0923+201 by adding a \heii\, component. The SFS is 
similar to that described in Section~\ref{sec:sfs} except for adding a Gaussian to account for the \heii\, 
line (see Figure~\ref{fig:fit-heii}).  In each individual spectrum, the \heii\, is too weak and highly 
blended with the \feii. Therefore, to reduce the degeneracy, the line width and shift of the \heii\, are 
fixed to the best values obtained by fitting the rms spectrum (in the bottom panel of Figure~\ref{fig:fit}). 
The measured light curves and time lag analysis are shown in Figure~\ref{fig:lc-heii}. The obtained time 
delays
are summarized in Table~\ref{table:heiilag}.

\begin{deluxetable}{lcccccc}
\renewcommand\arraystretch{1.4}
\centering
\tabletypesize{\scriptsize}
    \tablecaption{\cblue Rest-frame time lag measurements by including a \heii\, component in 
    spectral decompositions \label{table:heiilag}}
    \tabletypesize{\footnotesize}
    \tablehead{
    \multirow{2}{*}{Object} &
		\multirow{2}{*}{$\rm Line$} &
		\multirow{2}{*}{$r_{\rm max}$} &
		\colhead{} &
		\multicolumn{1}{c}{$\tau_{\rm cent}$} &
		\multicolumn{1}{c}{$\tau_{\rm MICA}$} & 
		\multicolumn{1}{c}{$\tau_{\rm JAV}$}\\
	\cline{5-7}
	\colhead{} &
	\colhead{} &
	\colhead{} &&
	\multicolumn{3}{c}{(day)}
    }
    \startdata
PG~0923+201  & ${\rm H\beta}$ & 0.92&& $110.1_{-10.5}^{+7.7}$ & $107.4 _{-3.9}^{+4.5}$ & $107.8_{-3.4}^{+3.0}$ \\
 & ${\rm H\gamma}$ & 0.93 && $111.9_{-7.6}^{+10.2}$ & $107.6 _{-4.4}^{+4.3}$  & $108.3 _{-3.8}^{+3.9}$\\
\enddata
\end{deluxetable}

\begin{figure*}[th!]
\centering 
\includegraphics[width=0.45\textwidth]{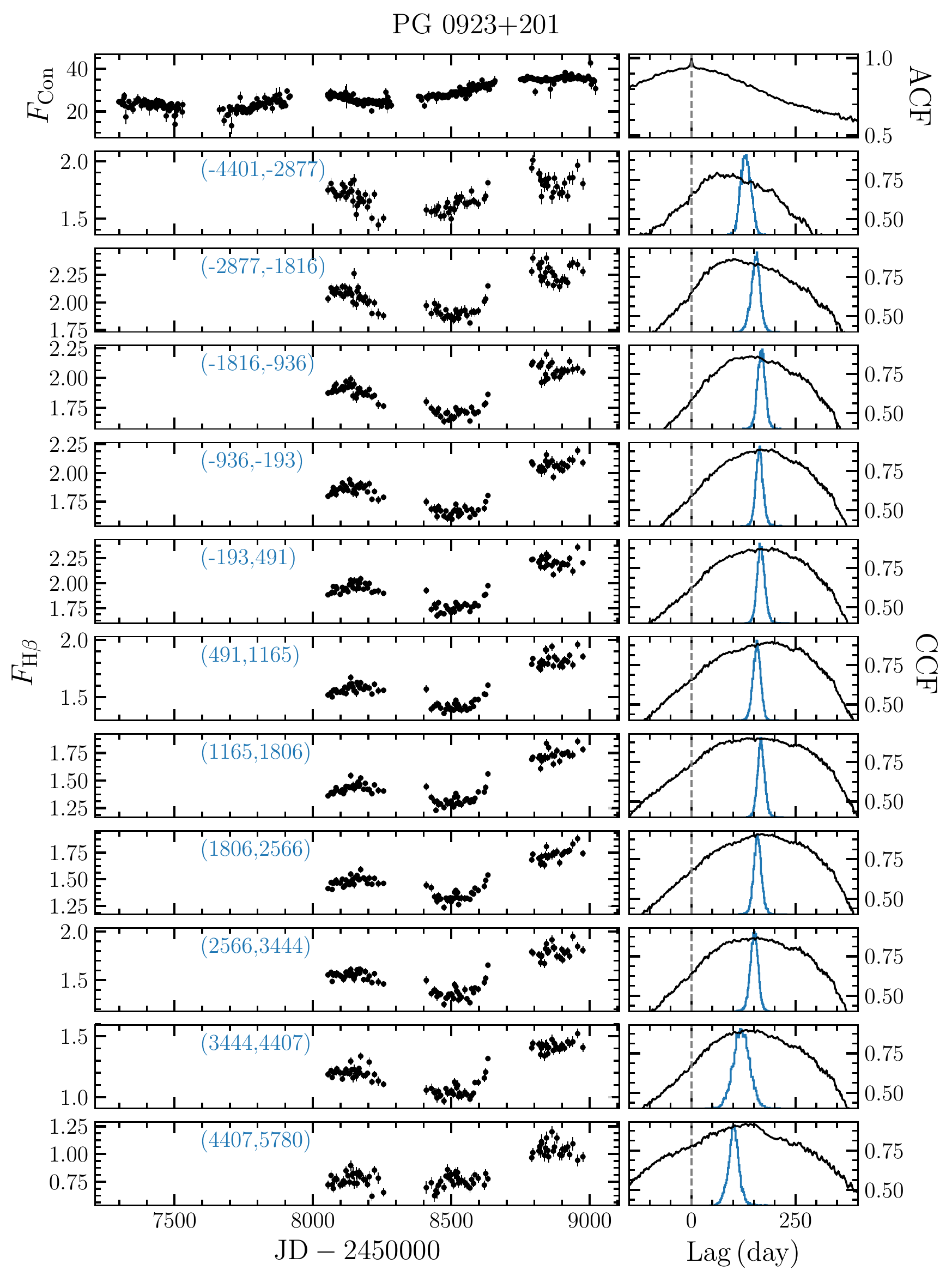}
\includegraphics[width=0.45\textwidth]{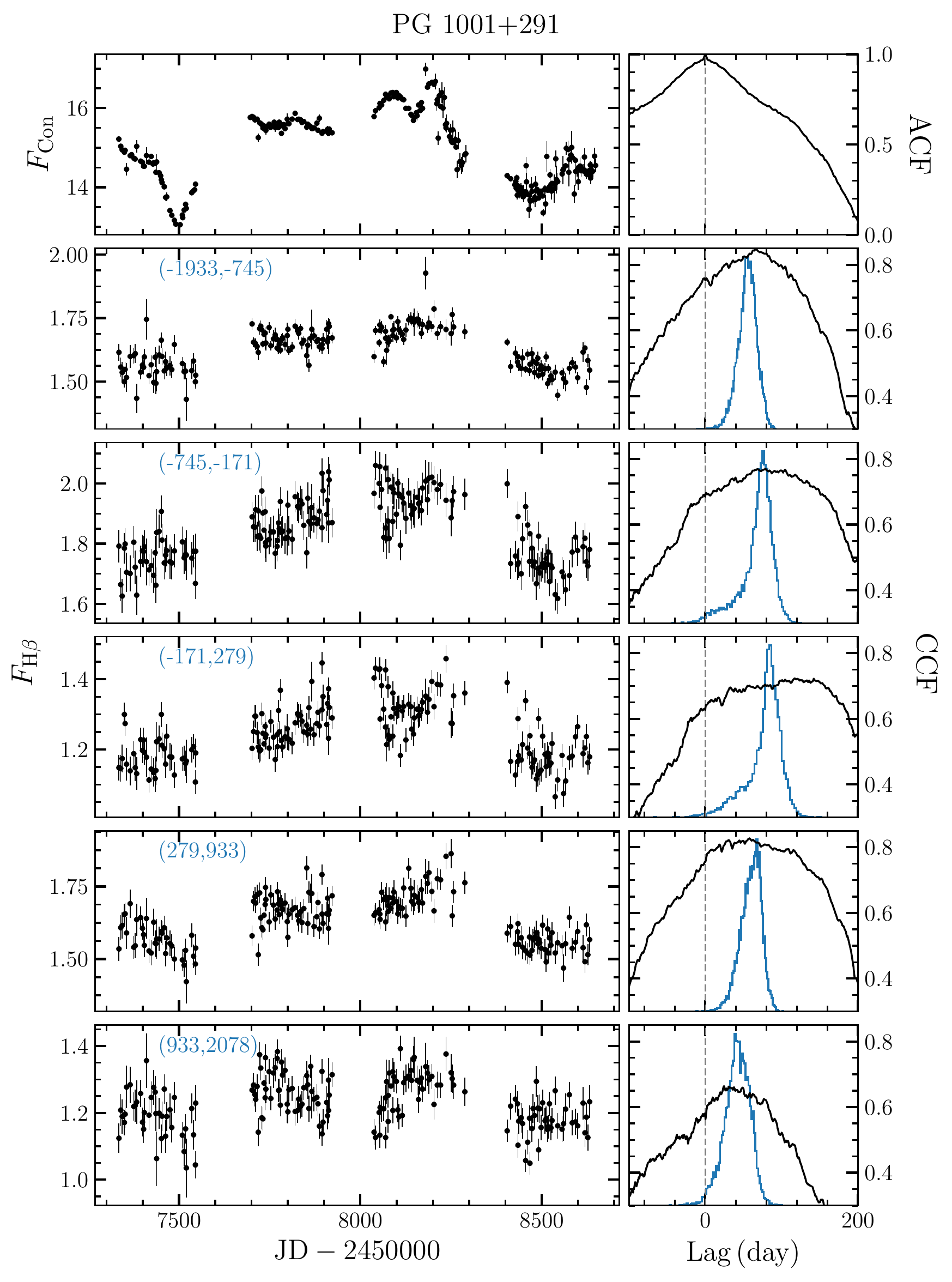}
\caption{The topmost panels show the continuum light curves and ACFs. The rest panels show the H$\beta$ 
light curves (black points with error bars), ICCFs (black lines) and CCCDs (blue histograms) at each velocity 
bin for (left) PG~0923+201 and (right) PG~1001+291. The units of the continuum and H$\beta$ light curves are
$10^{-16}~\ergscm$~\AA$^{-1}$ and $10^{-14}~\ergscm$, respectively. The velocity bin edges are marked by pairs 
of numbers (in units of km~s$^{-1}$). }
\label{fig:binlc}
\end{figure*}

{\cblue{
\section{ICCF Results in Each Velocity Bin}
}}
In Figure~\ref{fig:binlc}, we show the H$\beta$ light curves, ICCFs and CCCDs at different velocity bins for PG~0923+201 and PG~1001+291. 

\end{document}